\documentclass[fleqn,usenatbib]{mnras}
\usepackage{newtxtext,newtxmath}
\usepackage[T1]{fontenc}
\usepackage{ae,aecompl}

\newcommand{\sub}[1]{_{\rm #1}}
\renewcommand{\sup}[1]{^{\rm #1}}
\newcommand{\vinf}{v_\infty}
\newcommand{\vimp}{v\sub{imp}}
\newcommand{\vearth}{v\sub{\oplus}}
\newcommand{\vesc}{v\sub{esc}}
\renewcommand{\deg}{^\circ}
\newcommand{\beq}{\begin{equation}}
\newcommand{\eeq}{\end{equation}}
\newcommand{\vvec}{\vec{v}}

\newcommand{\hl}[1]{{\color{black} #1}}
\newcommand{\hll}[1]{{\color{black} #1}}
\newcommand{\hlll}[1]{{\color{black} #1}}

\usepackage{graphicx}	
\usepackage{amsmath}	
\usepackage{amssymb}	

\usepackage{algorithm}
\usepackage[noend]{algpseudocode}
\usepackage{hyperref}
\usepackage{color}

\title[Theoretical Distribution of Impacts]{Towards a theoretical determination of the geographical probability distribution of meteoroid impacts on Earth}

\author[Jorge I. Zuluaga \& Mario Sucerquia]{
Jorge I. Zuluaga\thanks{E-mail: jorge.zuluaga@udea.edu.co} and
Mario Sucerquia.\thanks{E-mail: mario.sucerquia@udea.edu.co}\\
Solar, Earth and Planetary Physics Group (SEAP) / FACom\\ 
Aerospace Science and Technology Research (ASTRA).\\
Instituto de F\'{\i}sica - FCEN, Universidad de Antioquia Calle 70 No. 52-21, Medell\'in, Colombia\\
}

\date{Accepted XXX. Received YYY; in original form ZZZ}
\pubyear{2017}

\begin{document}
\label{firstpage}
\pagerange{\pageref{firstpage}--\pageref{lastpage}}
\maketitle

\begin{abstract}
\hl{Tunguska and Chelyabinsk impact events occurred inside a geographical area of only 3.4\% of the Earth's surface. Although two events hardly constitute a statistically \hlll{significant} demonstration of a geographical pattern of impacts, their spatial coincidence is at least tantalizing. To understand if this concurrence reflects an underlying geographical and/or temporal pattern, we must aim at predicting the spatio-temporal distribution of meteoroid impacts on Earth. For this purpose we designed, implemented and tested a novel numerical technique, the ``Gravitational Ray Tracing'' (GRT) designed to compute the relative impact probability (RIP) on the surface of any planet. GRT is inspired by the so-called ray-casting techniques used to render realistic images of complex 3D scenes. In this paper we describe the method and the results of testing it at the time of large impact events. Our findings suggest a non-trivial pattern of impact probabilities at any given time on Earth. Locations at $60-90\deg$ from the apex are more prone to impacts, especially at midnight. Counterintuitively, sites close to apex direction \hlll{have} the lowest RIP, while in the antapex RIP are slightly larger than average. We present here preliminary maps of RIP at the time of Tunguska and Chelyabinsk events and found no evidence of a spatial or temporal pattern, suggesting that their coincidence was fortuitous. We apply the GRT method to compute theoretical RIP at the location and time of 394 large fireballs. Although the predicted spatio-temporal impact distribution matches marginally the observed events, we successfully predict their impact speed distribution.}
\end{abstract}

\begin{keywords}
methods: numerical -- meteorites, meteors, meteoroids.
\end{keywords}



\section{Introduction}

\hl{The surfaces of the Earth, the Moon and other solid objects in the Solar System have been systematically impacted by asteroids and meteoroids during the whole history of the planetary system.} Nowadays, the flux of impacts on those bodies have been greatly reduced. On Earth, however, even the impact of small objects (\hll{several meters to tens of meters across}) poses a severe threat to the more complex forms of life: human societies. Any effort intended to understand the flux and distribution of those impacts is critical for assessing the stability and even survival of our civilizations.  

In the last century, humans witnessed two of the largest impacts in recent history: the Tunguska and Chelyabinsk events (hereafter C-T events)\footnote{A third probable large event ($E\sim$1.5 Mt) happened on 3 august 1963 off the coast of Africa \citep{Revelle1997,Silber2009}. However, no independent confirmation of the event is available.}.  Unlike any other large impacts of the preceding centuries (that probably would have happened far off the coasts), those events happened over continental areas, had eye-witnesses and caused human injuries and damages. More interestingly the two events happened \hl{just 2,360 km apart}. 

Assuming a uniform geographical distribution of impacts, the probability that two unrelated events happen that close is 3.4$\%$ (the ratio between the area of a spherical cap with a radius of 2,400 km and the total area of the Earth). This mean that $\sim 30$ impacts (1/0.034) should occur on average to have two of them falling in a geographical area of a similar size.  Modern estimates of the rate of impacts  \citep{Silber2009,Brown2013} show us that fireballs with energies $E$ larger than that of Chelyabinsk event, namely $E>0.5$ MT, would happen with a frequency of 0.007 \citep{Harris2012} to 0.05 per year \citep{Revelle1997,Brown2013}.  Therefore, the average time required to have a geographical coincidence similar to that observed with the C-T events is 600-4,200 years.  Equivalently, the probability that two large impacts happened as close as C-T events in one century, is $2\%-14\%$.

Although these simple estimations do not demonstrate the existence of a geographical pattern of impacts around the Tunguska and Chelyabinsk regions, it is at least interesting to ask if the probabilities of having impacts on certain geographical areas and at given times in the year are always the same (uniform spatio-temporal distribution) or if impact probability has a more complex behavior than this simple and common a priori assumption.  

The distribution of asteroids and meteoroids in the Solar System \hl{is not uniform, either in physical or configuration space} (see eg. \citealt{Bottke2002}, \citealt{Jeongahn2014}, \citealt{Granvik2016} and references therein). Although the Earth is a small planet as compared to the scale of asteroid orbits and one may expect that impactors should come from every direction in sky, the complex relative dynamics between the population of parent bodies and the Earth, the focusing effect of Earth's gravitational field and even the presence of the Moon, could create non-trivial geographical patterns of impacts.  

Other authors have extensively studied the problem of spatial and temporal asymmetries in the flux of impactors on the surfaces of the Moon, the Earth and other terrestrial planets in the Solar System   \citep{Halliday1964,Wetherill1968,Wetherill1985,Halliday1982,Morbidelli1998,LeFeuvre2005,LeFeuvre2008,Gallant2009, Werner2010, Ito2010}. They have mainly focused on describing latitudinal asymmetries, non-uniform crater distribution, relative polar to equatorial impact fluxes, seasonal and day/night asymmetries, among other non-trivial effects. These works have conclusively demonstrated that, at least for the case of the Earth and the Moon, the average distributions of impacts is not uniform and depends on latitude, season and time of the day. Still, the aim of those works has not been assessing the risk of meteoroid impact at a given time or location on the Earth, which could be key information for present and future efforts intended to evaluate the threats posed by asteroid impacts.

In this paper we present a novel method to study the instantaneous geographic distribution of meteoroid impacts on a given planetary body in the Solar System.  The method is particularly well suited for assessing the risk of impact on specific geographical areas on Earth.  We  call the new method, ``Gravitational Ray Tracing'' (hereafter GRT). \hl{GRT} is inspired by \hl{the} ray-casting algorithms, commonly used in computer graphics to render complex visual scenes (see eg. \citealt{Goldstein1971, Roth1982, Weghorst1984, Comninos2010}).

\hl{This work can contribute to the understanding} of the spatio-temporal distribution of impacts on Solar System's objects in several new ways: (1) providing an efficient new approach for the calculation of impact probabilities; (2) testing a backward-integration method that can be applied to a broader range of problems than the traditional forward-integration techniques;  and last but not least (3)  computing impact probability at specific locations on Earth (cities, countries or broader geographical areas) that could be used for impact risk assessment.

It is important to stress that although our initial motivation was to solve the tantalizing spatial coincidence between the C-T events, we do not aim in this paper at probing the existence of a relationship between both impact events. \hl{Neither are we} using their apparent coincidence as a working hypothesis for developing our methods. Our ultimate goal here is to design, implement and test a method that can be used to study this and many other related problems.

This paper is organized as follow: in \autoref{sec:GRT} we describe in detail the GRT method. \autoref{sec:results} is devoted to testing the method for a specific set of cases. In \autoref{sec:discussion} we discuss the implications of the results, the limitation of the method and the future prospects of its application. Finally in \autoref{sec:conclusions} we summarize our results and draw the conclusions of this work. 


\section{Gravitational Ray Tracing}
\label{sec:GRT}

Several analytic and numerical techniques have been devised to calculate spatio-temporal anisotropies in the flux of meteoroids or asteroids on the surface of Solar System bodies (see eg. \hl{ \citealt{LeFeuvre2008,Gallant2009,LeFeuvre2011,Wang2016,zuluaga2017p}} and references therein).  Most of these techniques are inspired or based on Monte Carlo methods  and use a ``forward-integration'' approach. Hundreds of thousands and even millions of test particles are thrown from interplanetary space towards a target object (eg. the Earth).  The synthetic population of impactors have orbital elements $q$, $e$, $i$, $\Omega$ and $\omega$ (with $q$ the perihelion distance, $e$ the eccentricity, $i$ the orbital inclination, $\Omega$ the longitude of ascending node and $\omega$ the argument of perihelion)  following a given distribution.  Using this distribution, initial conditions for test particles in physical space, $x$, $y$, $z$, $v_\mathrm{x}$, $v_\mathrm{y}$ and $v_\mathrm{z}$ are generated near to the target object. Their trajectories are then integrated forward in time until they impact or miss the target. Once a significant number of test particles achieve to  hit the object the statistical properties of impacts are computed.

Although forward-integration schemes have demonstrated their value at estimating  impact probability distribution, these methods are inefficient (i.e. the number of particles impacting the target body is much smaller than the total number of particles used) and, for a limited amount of computational resources (number of CPUs and hard-disk space), they are not able to achieve enough resolution to asses general impact risk on specific locations on Earth. 

\begin{figure*}  
  \centering
  \vspace{0.2cm}
   \includegraphics[scale=0.7]{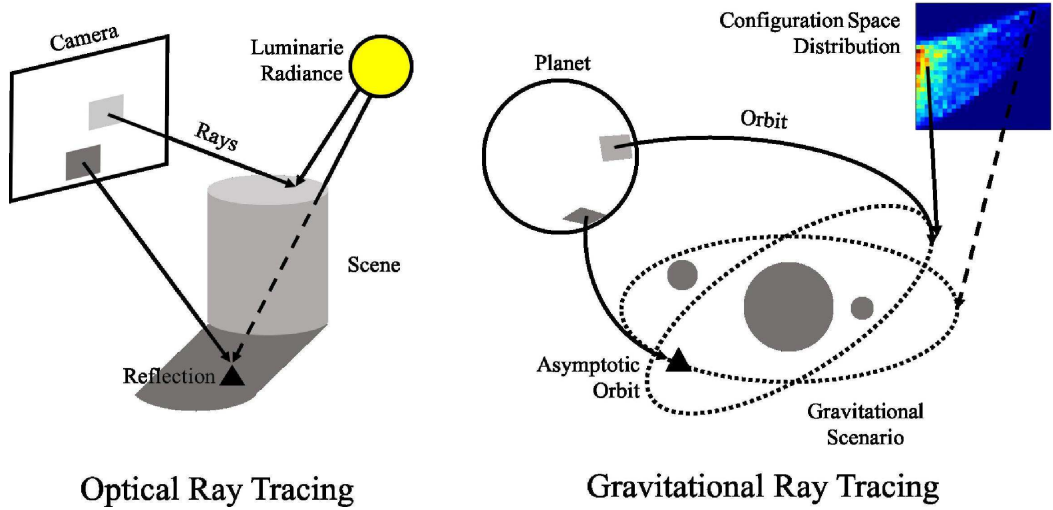}\\\vspace{0.5cm}
  \scriptsize
  \caption{Schematic representation of the optical and gravitational ray-tracing ( see text and Algorithm \ref{alg:GRT} for a detailed description). The ``gravitational scenario'' stands for the bodies involved in the force field calculation.  The ``asymptotic orbit'' is the orbit that the test particle reaches when its trajectory is integrated backwards.}
\label{fig:GRT}
\end{figure*}

How can these traditional Monte Carlo methods be improved?. An optical analogy could help us to find an answer (see \autoref{fig:GRT}).  

If instead of computing the impact probability distribution over the surface of a planetary body, we would want to calculate the intensity of the light received in the focal plane of a camera when exposed to light coming form a scene (light intensity here is the analogue to impact probability and the camera is  the analogue of the target body surface), a forward-integration technique would proceed this way: photons (asteroids and meteoroids) would be thrown from the light sources towards the objects in the scene. The radiance (distribution of asteroids in configuration space) of the luminaries must be well known. Observed intensity (impact probability)  would be computed after following the trajectories (orbits) of all photons, while bouncing off or traversing the objects in the complex scene (gravitational field of the Solar System).  Many photons would hit the camera, but most of them would be lost when they are either absorbed by dark objects (collide against astronomical bodies) or  abandon the scene and never reach the focal plane of the camera (escape from the Solar System).

A more efficient way to perform this calculation in the optical problem was first described in the seminal papers by \citet{Goldstein1971} and \citet{Roth1982}. The  devised technique is called ``ray-tracing'' or ``ray-casting''. We implement in the case of meteoroid and asteroid impact in Solar System bodies, the key ideas of these methods. The resulting technique will be conveniently called ``Gravitational Ray Tracing'' (GRT).

\autoref{fig:GRT} and Algorithm \ref{alg:GRT} explain schematically the GRT method, that can be described briefly as follows. In the optical (gravitational) ray-tracing, the path of photons (asteroids and meteoroids) are integrated backward from the camera (Earth's surface) towards and through the objects in the scene (Solar System gravitational field). Once there, the intensity (impact probability) associated with each ray (meteoroid incoming trajectory) is computed by calculating the flux (density in configuration space) of the light coming from the luminaries.

\begin{algorithm}
\small
\caption{Gravitational Ray Tracing algorithm
  \label{alg:GRT}}
  \begin{algorithmic}[1]
  \State Set: \textit{target object, date, time.}
  \State Set: Integration time, $T_{\rm max}\sim$ \textit{target orbital period}.
  \State Generate: $M$ random geographical positions:  \textit{longitude}$[i]$, \textit{latitude}$[i]$.
  \ForAll{i}:
  \State \hl{Initialize probability: $P[i]\gets 0$ }
  \State Generate: $N$ random directions: \textit{azimuth}$[j]$, \textit{elevation}$[j]$.
  \ForAll{j}:
  \State Generate: $L$ random impact speeds: $v_{\rm imp}[k]$.
  \ForAll{k}:
  \State Integrate trajectory backward until (\textit{date,time}) - $T_{\rm max}$
  \If{Particle collide with an object}
  \State Continue to next $k$.
  \Else
  \State Get final \hl{state}: $x$, $y$, $z$, $v_{\rm x}$, $v_{\rm y}$, $v_{\rm z}$. 
  \State Calculate orbital elements: $q$, $e$, $i$,  $\Omega$, $\omega$
  \State \hl{Calculate progenitor's density: $R$($q,e,i,\Omega,\omega$)}
  \State \hl{ $P[i]\gets$ $P[i]$ + $R$(q,e,i,$\Omega,\omega$). }
  \EndIf
  \EndFor 
  \EndFor 
  \State Normalize site probability: \hl{ $P[i]\gets$ Norm. $\times$ $P[i]$}
  \EndFor
  \end{algorithmic}
\end{algorithm}

 The usage of backward-integration methods for the study of planetary impacts in the Solar System is not new. As a matter of fact, backward-integration is the only suitable method to reconstruct the trajectories of meteor and bolide parent bodies (see eg. \citealt{Dmitriev2015} and \citealt{Zuluaga2013}). The method has been also used in the past for similar purposes than those pursued in this work (see eg. \citealt{Takikawa1989,Tanikawa1991}).  The implementation of backward-integration in GRT, however, is novel in the sense of implementing many of the lessons learned with its optical counterpart.

In the following paragraphs we will describe each component of the GRT algorithm.

\subsection{Initial Conditions}
\label{subsec:initial}

Generation of proper initial conditions is  critical for the ray-tracing algorithm. The camera focal plane (planetary surface) and the directions in which photons (meteoroids) are thrown must be sampled properly to avoid inefficiencies and numerical artifacts. The most common \hl{artifact} is called spatial and temporal aliasing. Aliasing arises when samples (positions and directions) are regularly spaced \citep{Dippe1985}. In our case aliasing could arise if geographic locations and impact directions are taken from a uniform grid of longitudes (azimuths) and latitudes (zenith angles).

\begin{figure*}  
  \centering
  \vspace{0.2cm}
   \includegraphics[scale=0.65]{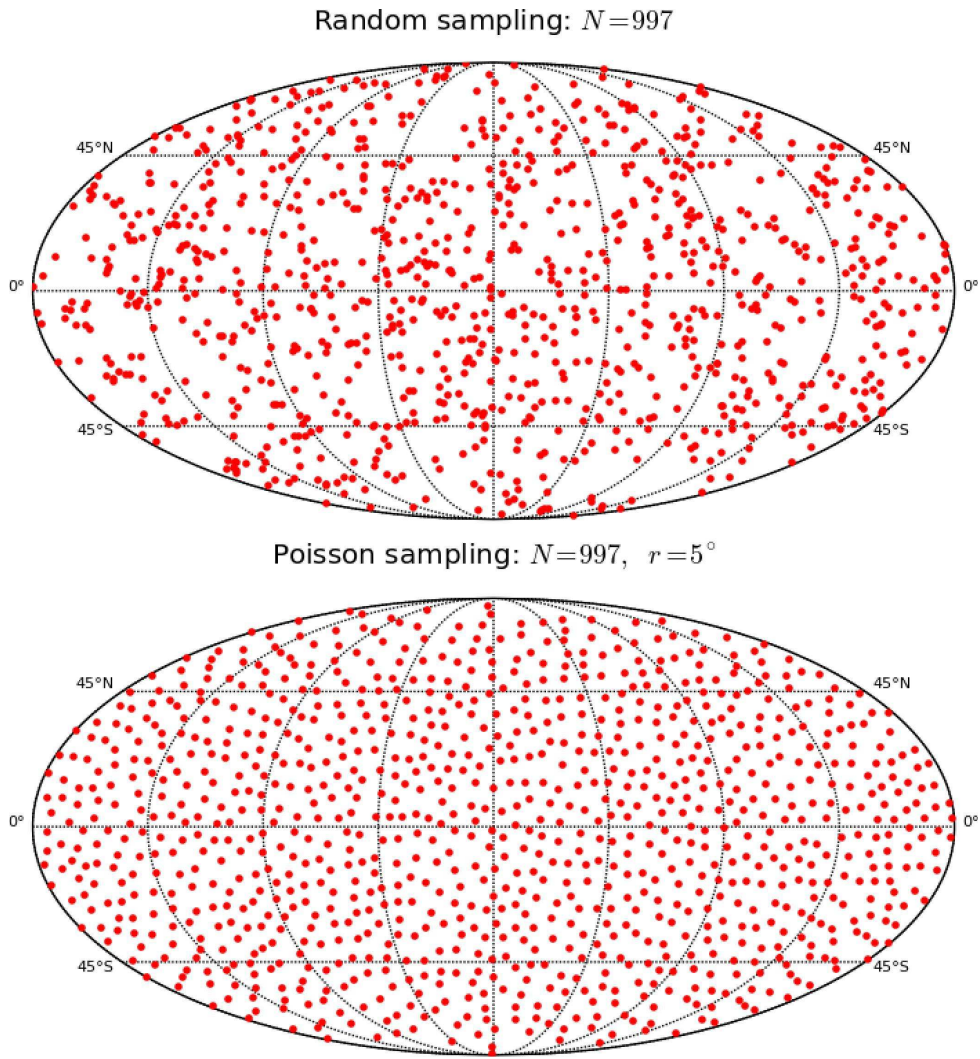}\\\vspace{0.5cm}
  \scriptsize
  \caption{Top panel: Random geographic positions generated with a traditional algorithm. Bottom panel: positions generated using a Poisson sampling algorithm.  The minimum  {angular} distance  {$r$} among points is 5$^\circ$. $1003$ points were finally accepted by the algorithm.}
\label{fig:GeographicPositions}
\end{figure*}

Generating random points on the surface of a unitary sphere is a common approach to sampling directions (see \hll{top} panel in \autoref{fig:GeographicPositions}). This sampling method tends to over and under represent geographical areas or directions in the sky. Ray-casting techniques have found a clever way to avoid this effect \citep{Dippe1985,Cook1986}: instead of random sampling we can use ``Poisson sampling''.  

In Poisson sampling, random points are generated uniformly over the sampled surface (planetary surface or the sky). Points that fall closer than a critical distance from one of their neighbors are discarded from the sample. New points are generated and discarded with the same conditions. The process is repeated until no other point could be placed in the surface. The resulting distribution of points is called a Poisson disk distribution or simply ``blue noise''\footnote{Interestingly the idea of Poisson sampling was originally inspired by the distribution of photo receptors in the retina (see \citealt{Cook1986} and references therein), that could arise from evolution to avoid visual aliasing effects.}.   In \autoref{fig:GeographicPositions} we compare uniform random points on a sphere and points generated with a Poisson sampling algorithm. For GRT we use Poisson sampling to generate both, geographical positions and impact incoming directions.

Once geographic positions and directions in the sky are generated, we need to draw random impact speeds. In forward-integration schemes, impact speeds are part of the products of the simulations. In GRT, where integration is reversed, we require an a priori assumption about this distribution.

\hl{Speed distribution depends on the impactor mass range (see eg. \citealt{Wiegert2009}).}  Meteors and fireball networks may provide \hl{valuable} information on impact speed distributions (see eg. \citealt{Drolshagen2014}).  \hl{However, up to 50\% of the objects detected by those networks belong to major meteor showers, whose objects, with rare exceptions (eg. the Taurid meteor shower), are relatively small (tens of centimeters), at least for the purposes pursued here}. On the other hand, in most cases, the data from this networks is not freely available and only some processed products have been published \hl{up to date}.  Another potential source of information are the observational and theoretical studies of the properties of potential impactors on space vehicles \citep{Drolshagen2008,Taylor1995}. \hl{Again, the size of the particles considered by those studies are very small not to mention that} these models only provide an initial speed distribution, $\vinf$, and no information on direction, which is required to convert $\vinf$ into impact speed, $\vimp$.

 In order to avoid potential biases and numerical artifacts, we use the simplest (though not the most efficient) solution: speeds are drawn uniformly in the interval $[11.1,43.6]$ km/s, \hl{with respect to a geocentric rotating reference system}. \hl{In this reference system}, particles with speeds relative to the Earth's surface lower than 11.1 km/s (Earth's escape velocity) would never reach heliocentric orbits. On the other hand, particles with \hl{surface} speeds larger than \hl{43.6 ($=\sqrt{(11.1+0.46)^2+42^2}$, being 0.46 km/s the Earth's rotational speed at the equator)} will end up in unbound orbits around the Sun (heliocentric velocities larger than 72 km/s, the Solar escape velocity at the Earth's distance).  
 
 \hl{It could be more convenient to generate heliocentric velocities in the range 0-72 km/s and transform them to velocities with respect to the Earth's surface taking into the account the focusing effect of Earth's gravity and Earth's translation and rotation. We have verified with numerical experiments that using a non-uniform prior speed distribution with respect to the surface of the Earth (as resulting from the transformation from an heliocentric to an Earth's surface reference frame), does not change significantly our main conclusions. Thus, we will prefer the simplest method to generate impact speeds, namely do it with respect to the rotating geocentric reference frame (no transformation required).}

\subsection{Integration}
\label{subsec:integration}

In order to integrate the trajectory of a test particle, we need first to determine with precision the position and velocity with respect to the Solar System Barycenter (SSB) of the impact site.  We use for this purpose the SPICE toolkit of the NASA Navigation and Ancillary Information Facility (NAIF) \citep{Acton1996}.  

The initial positions of the test particles are calculated assuming that all impacts happen at the same height above sea level (eg. 80 km). Initial velocity with respect to SSB is  {calculated} using:

$$
\vvec\sub{ini}=M\sub{ECJ2000}(t)M\sub{loc}(t)(\vvec\sub{imp}+\vvec\sub{rot})+\vvec\sub{\oplus}
$$

where the initial \hll{$\vec{v}\sub{imp}$} and surface rotational velocity $\vec{v}_{\rm rot}$ are computed from:

\hll{
\begin{eqnarray}
\nonumber
\vvec\sub{imp} & : & (\vimp \cos A \sin z,-\vimp \sin A \sin z,\vimp \cos z)\\
\nonumber
\vvec\sub{rot} & : & (0,-2\pi\rho/P\sub{rot},0)
\end{eqnarray}
}

Here $A$ and $z$ are the azimuth and zenith angle of the impact radiant, $\rho$ is the distance from the impact site to the rotational axis of the body and $P\sub{rot}$ is the rotational period. $M\sub{loc}(t)$ and  $M\sub{ECJ2000}(t)$ are the instantaneous rotation matrices transforming from the local reference frame to the Earth fixed ITRF reference frame and from there to the inertial Ecliptic J2000 reference frame, respectively.

In GRT initial positions and velocities can be generated within few km and several m/s from the impact site (eg. a small city). This is one of the advantages of the method  {with} respect to forward-integrating methods  {for which the resolution of the final impactor position is limited by the precision of the integrator}. 

 {In all cases} we integrate the orbit from the impact time, back to 6 months before.  For that purpose we use a Gragg-Bulirsch-Stoer integrator \citep{Gragg1965,Bulirsch1966} adapted from\footnote{\url{http://www.mymathlib.com/}}. The use of a high-order, precise integrator allows us to take into account  {on one hand}, the interaction with close objects such as the Moon, and on the other hand, to evaluate collisions with these objects and  {with} the Earth itself.  We include in the force field all major planets.  Positions and velocities of the  {the planets were} not computed with the integrator itself, but taken directly from the DE430 SPICE kernel\footnote{\url{http://naif.jpl.nasa.gov/pub/naif}}.

We test the precision of the integrator by propagating the orbits of the Moon and Eros in the force field of the planets.  Initial conditions in both cases were calculated with the JPL's Horizon system\footnote{\url{http://ssd.jpl.nasa.gov/horizons.cgi}}.  The orbits of a fake Moon and a fake Eros were integrated both forward in time and forward-then-backward. When integrated forward in time we compare the resulting positions and velocities with those obtained from the SPICE kernel (that contains the actual position of the Moon and Eros). When integrated forward-then-backward we compare positions and velocities at the same times in both directions.  We find that our integrator has a maximum fractional precision of the order of $10^{-8}$ that for a typical NEO corresponds to an absolute precision below 1 km which is, as explained before, very convenient for achieving spatial resolutions at the scale of small cities.

\subsection{NEOs distribution}
\label{subsec:source}

\begin{figure*}  
  \centering
  \vspace{0.2cm}
\includegraphics[width=56mm]{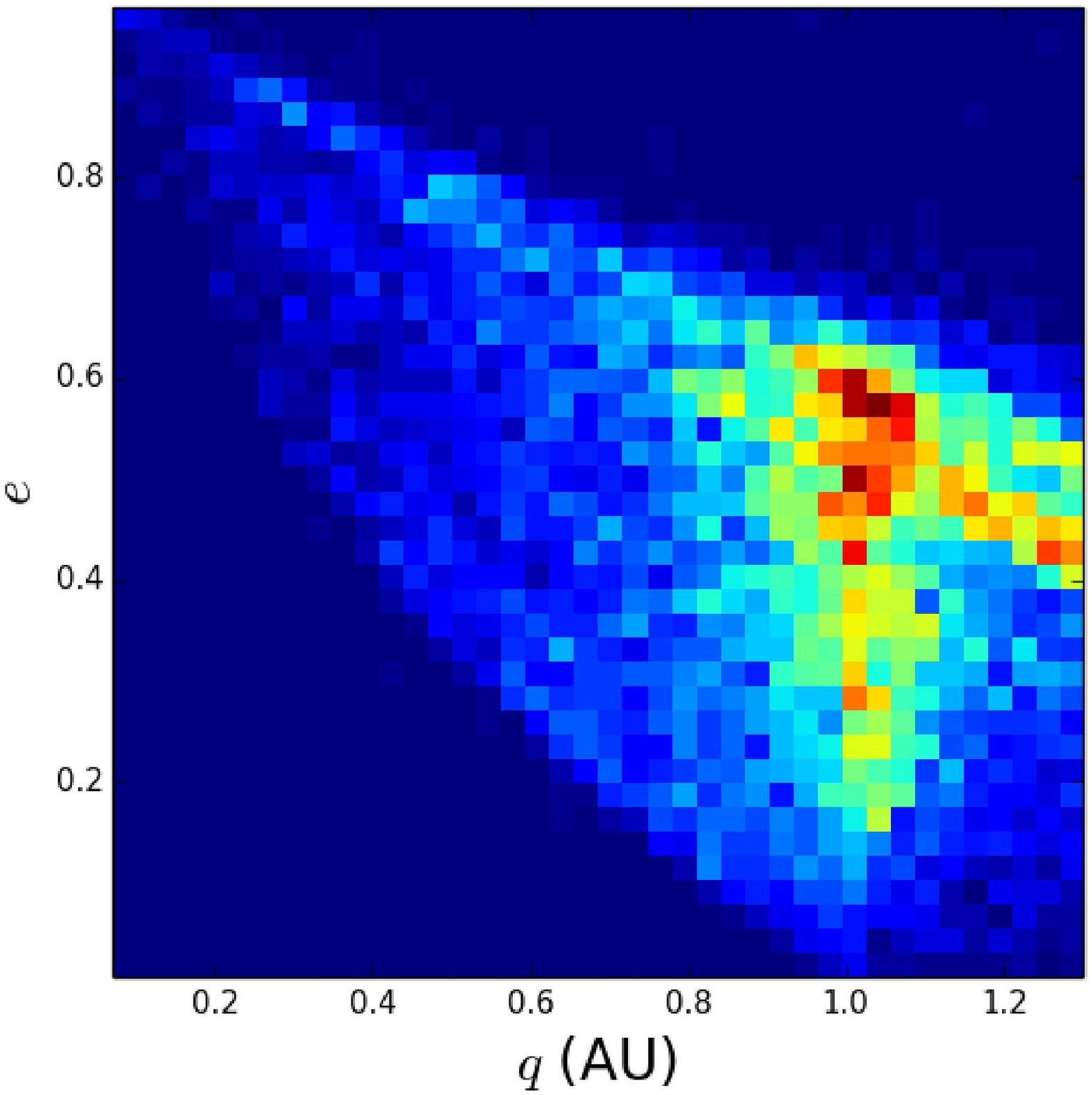}
\includegraphics[width=58mm]{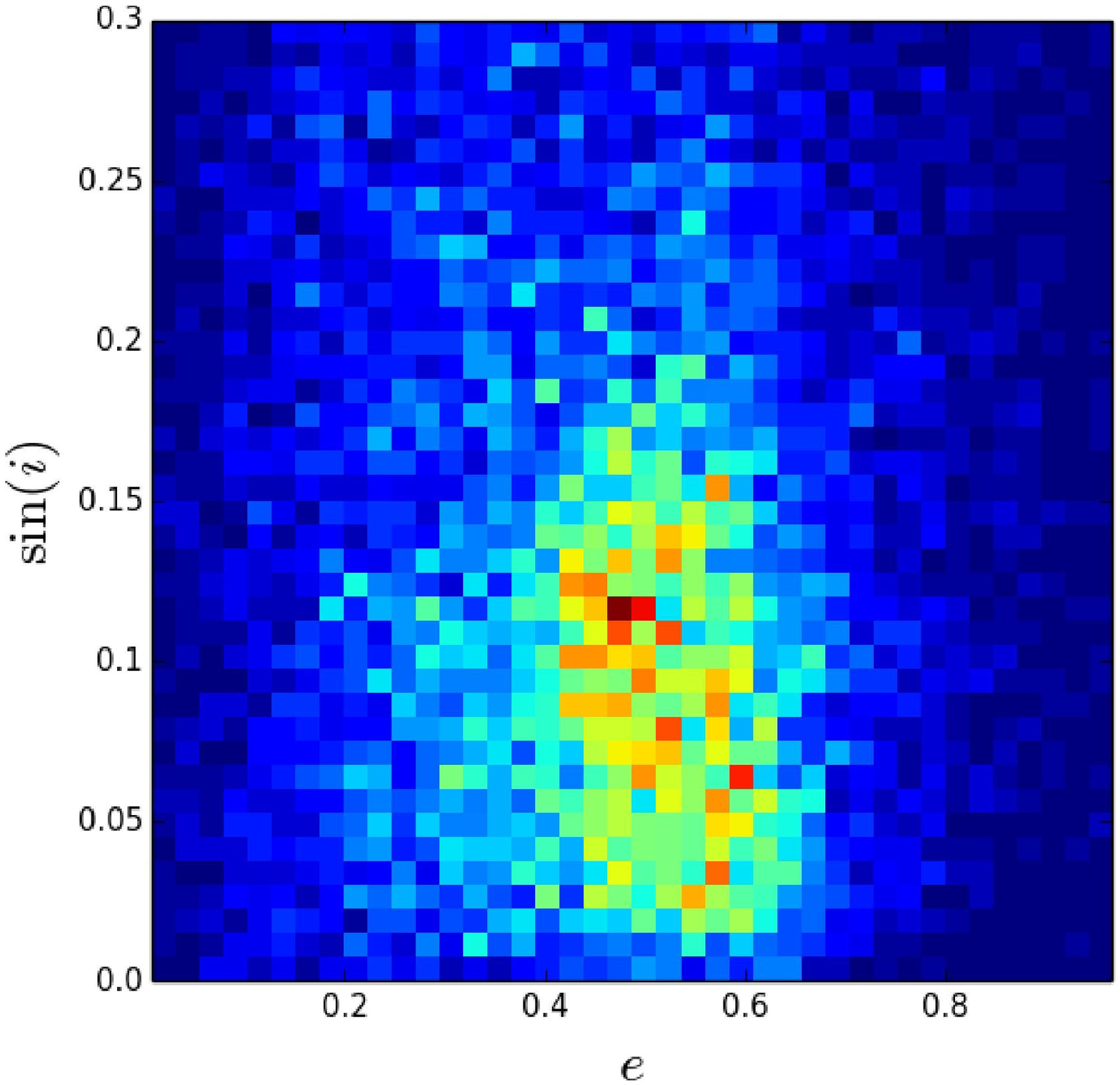}
\includegraphics[width=57mm]{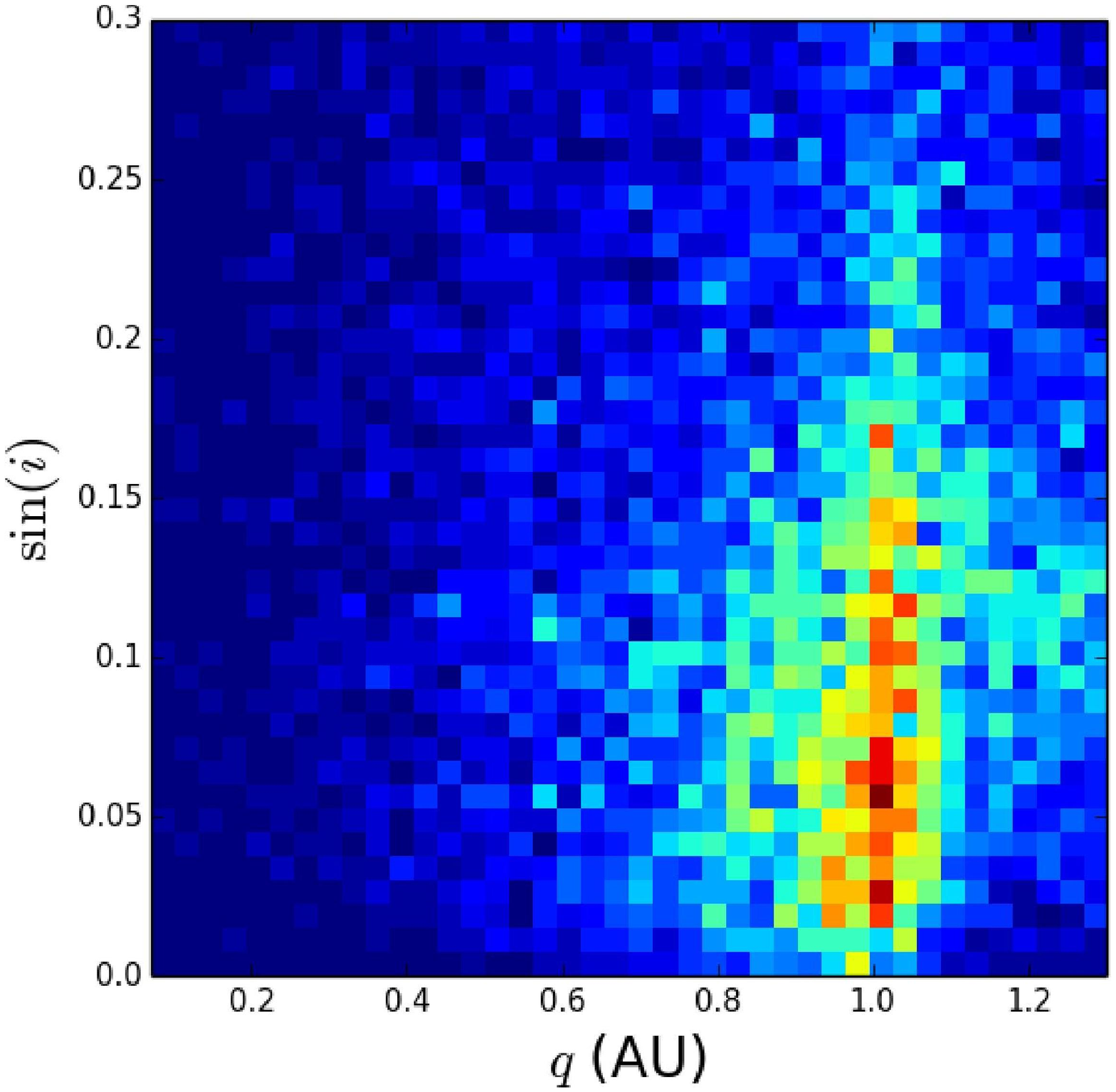}
\\\vspace{0.5cm}
\includegraphics[width=58mm]{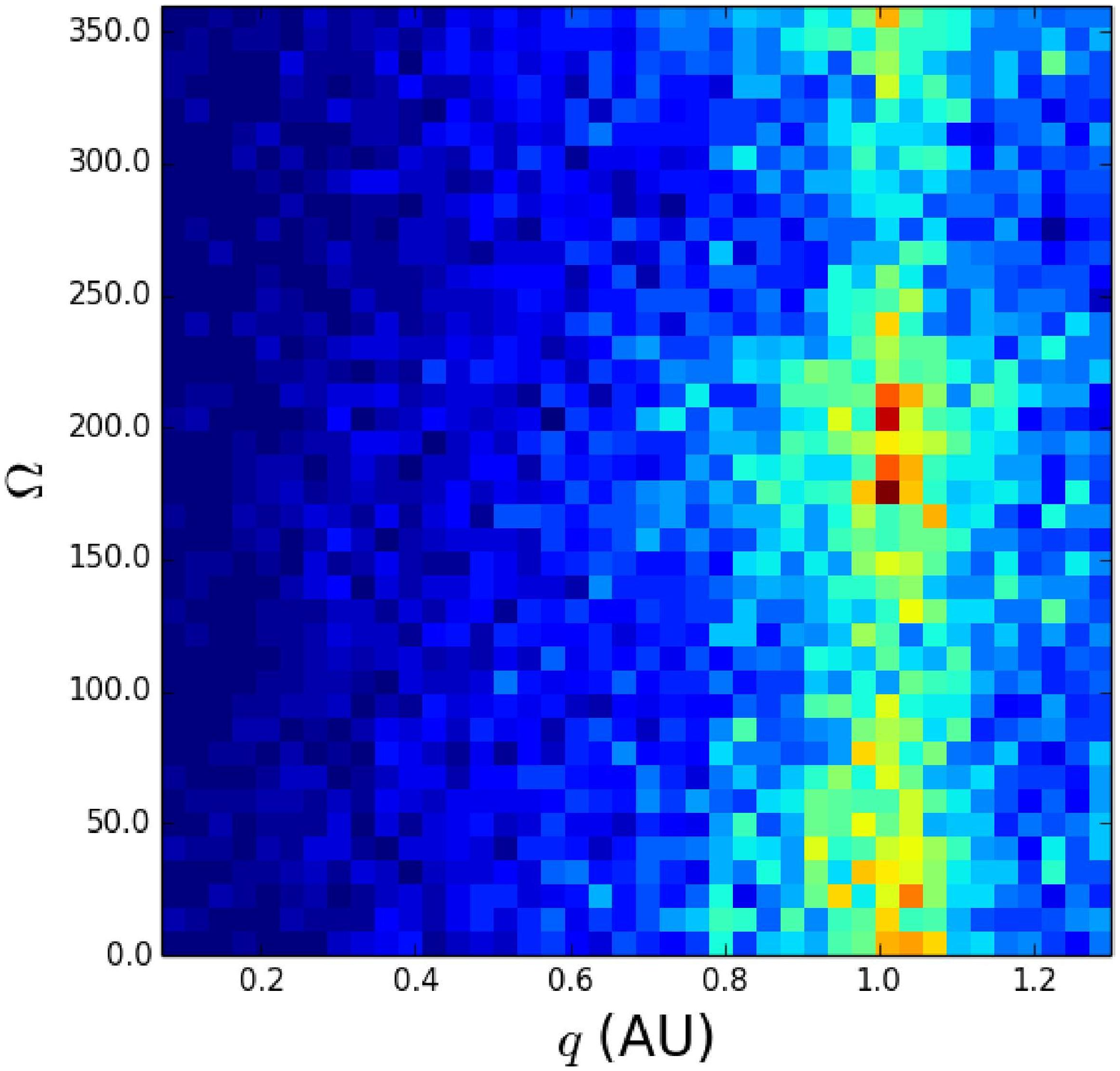}
\includegraphics[width=58mm]{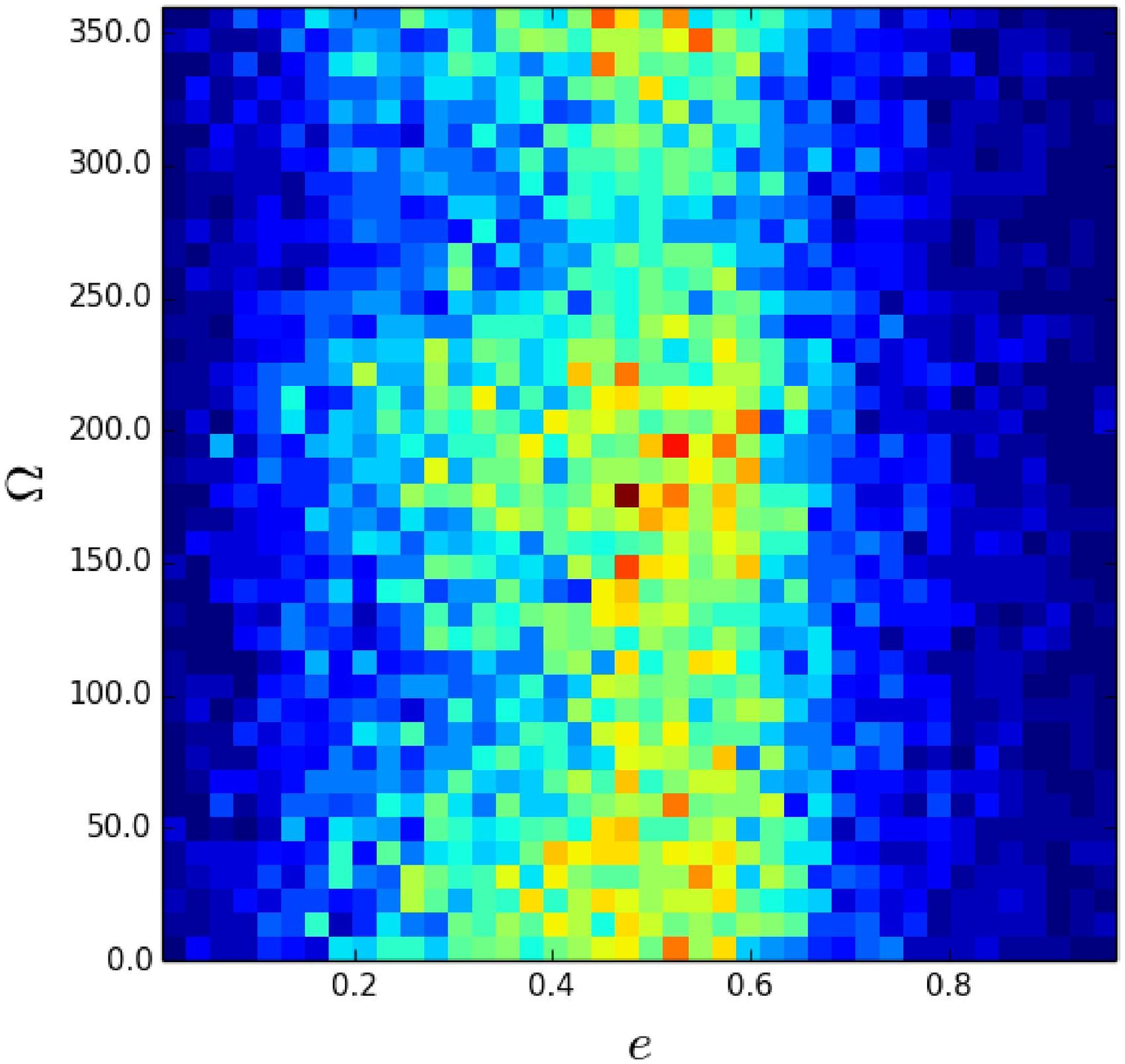}
\includegraphics[width=58mm]{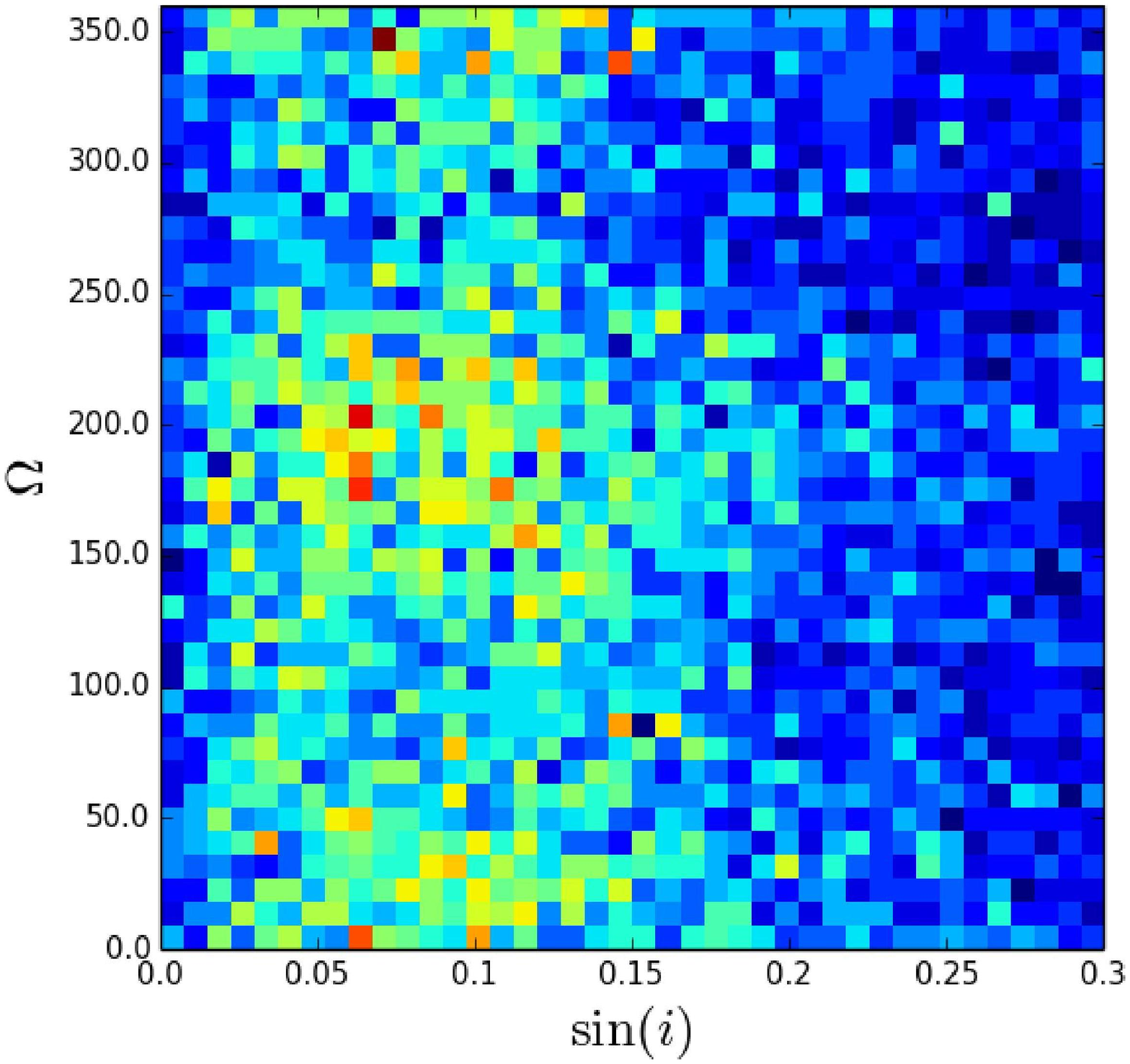}
\\\vspace{0.5cm}
\includegraphics[width=58mm]{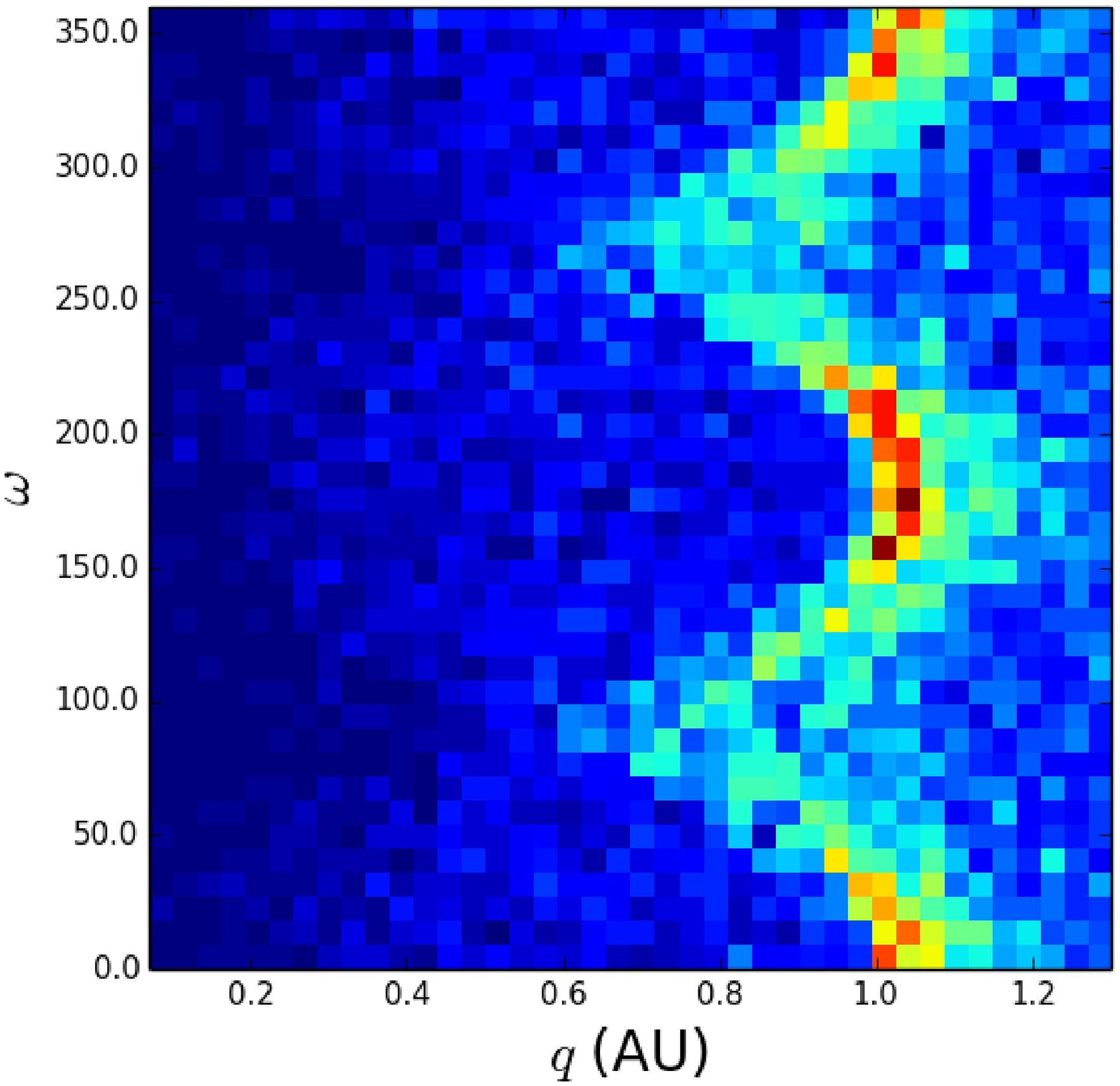}
\includegraphics[width=58mm]{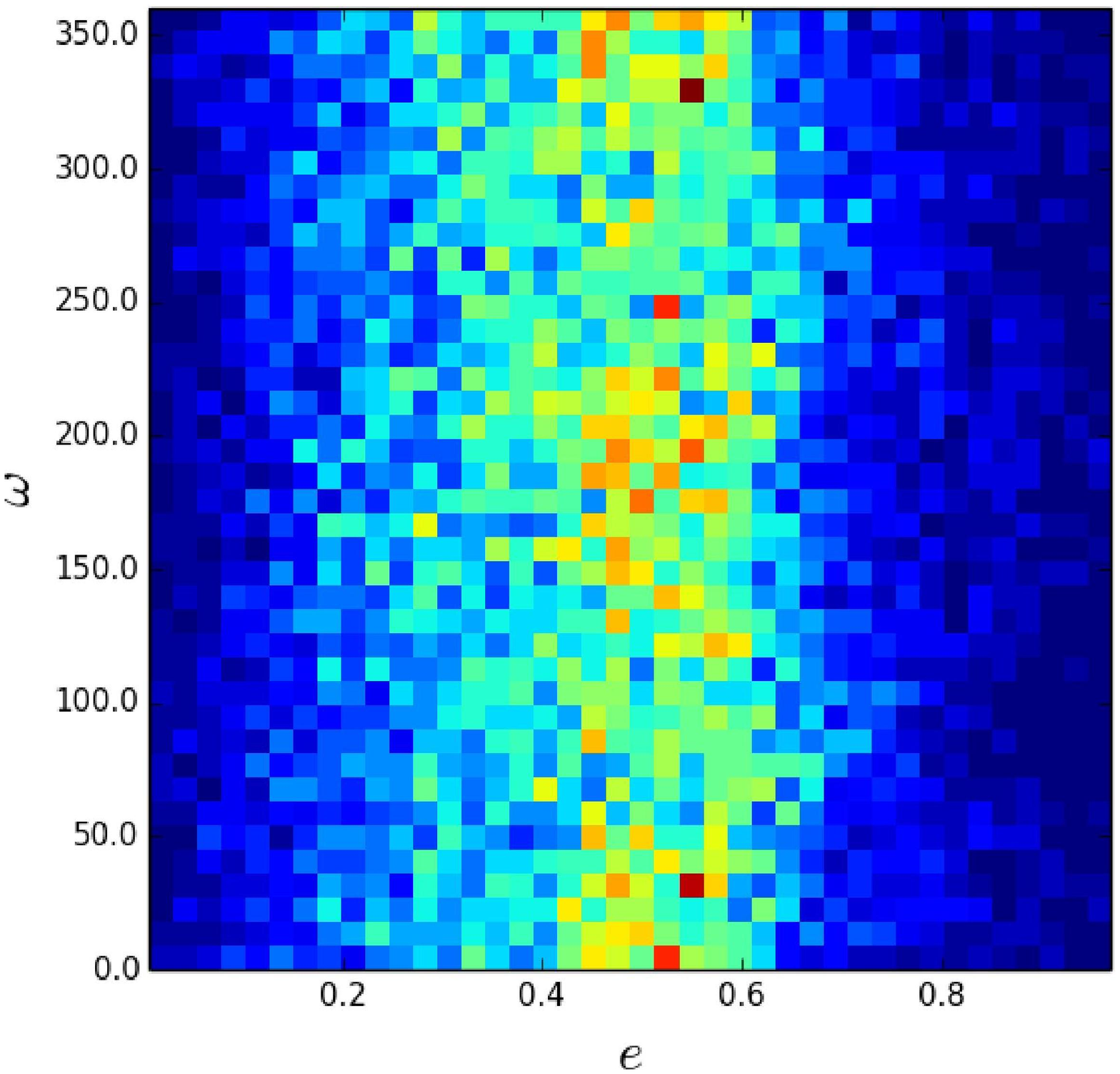}
\includegraphics[width=58mm]{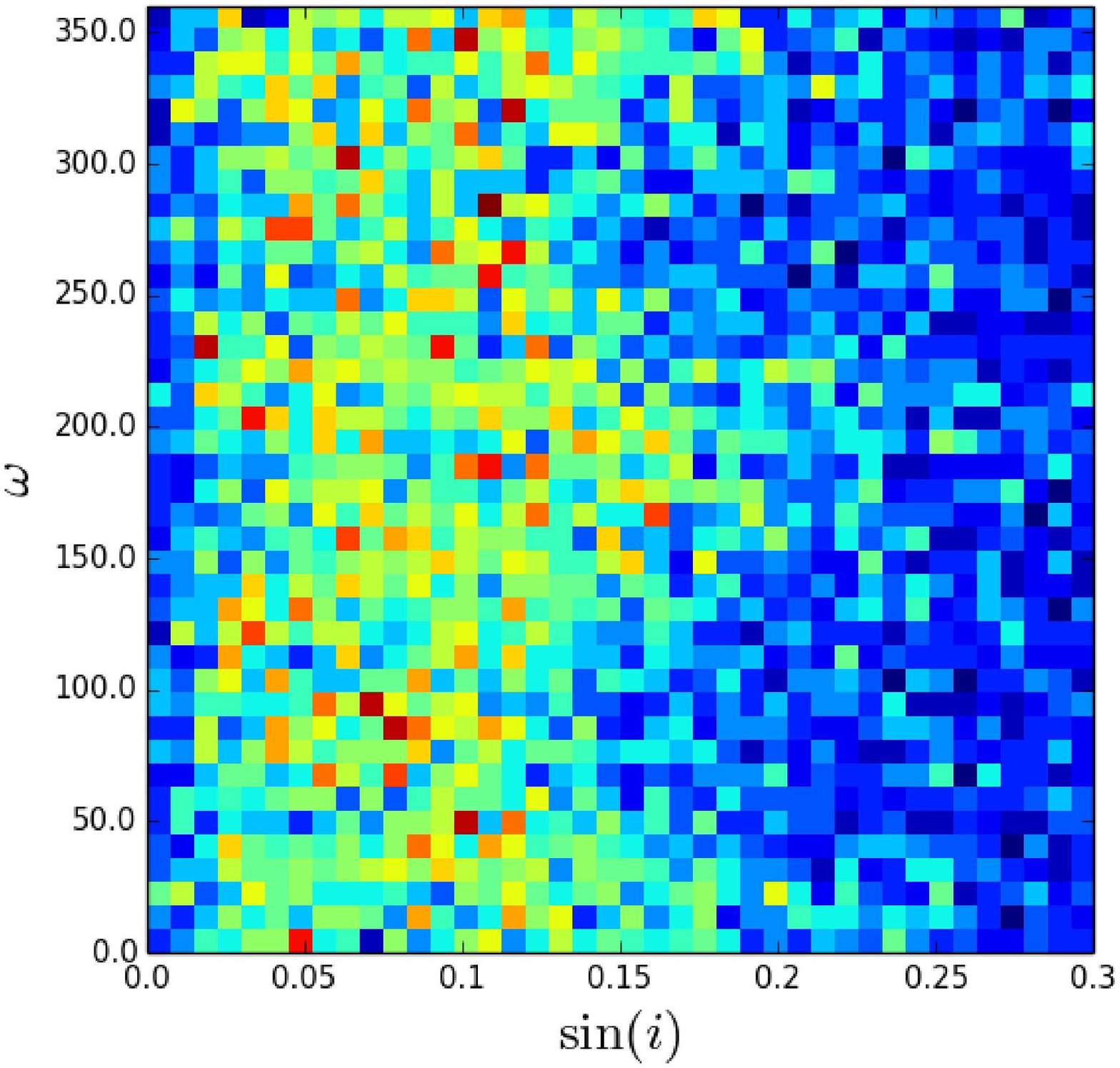}
\\\vspace{0.5cm}
   \scriptsize
  \caption{\hl{Schematic} density diagrams of the orbital properties for the complete sample of 14,291 NEOs. The color of each point in the diagrams represent the number of objects that have properties similar to the \hll{corresponding} coordinates irrespective of the value of the other orbital properties (marginal two-dimensional distributions). \hll{Since density could be very different from diagram to diagram, the colors in each panel are not comparable (they represent different densities)}}.\vspace{0.2cm}
\label{fig:NEODistribution}
\end{figure*}

To calculate impact probability we need the density of parent bodies in configuration space, ($q$,$e$,$i$,$\Omega$,$\omega$).  In \autoref{fig:NEODistribution} we represent the density of the orbital properties for 14,291 objects cataloged as NEOs in the NASA Small Body database\footnote{As obtained from the JPL Small-Body Database Search Engine \url{http://ssd.jpl.nasa.gov/sbdb_query.cgi} in September 2016}.

\begin{figure*}  
  \centering
  \vspace{0.2cm}
\includegraphics[width=56mm]{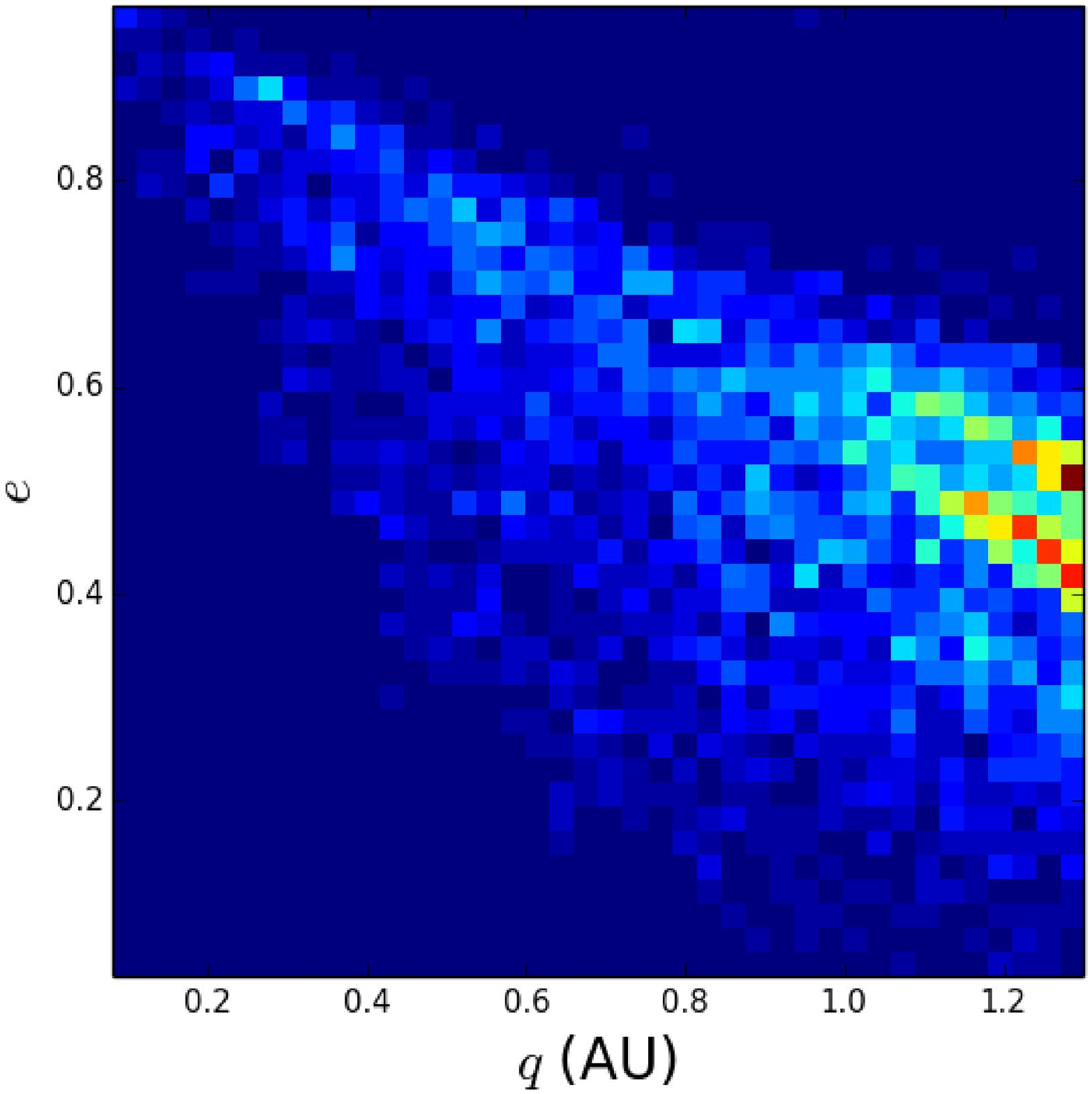}
\includegraphics[width=58mm]{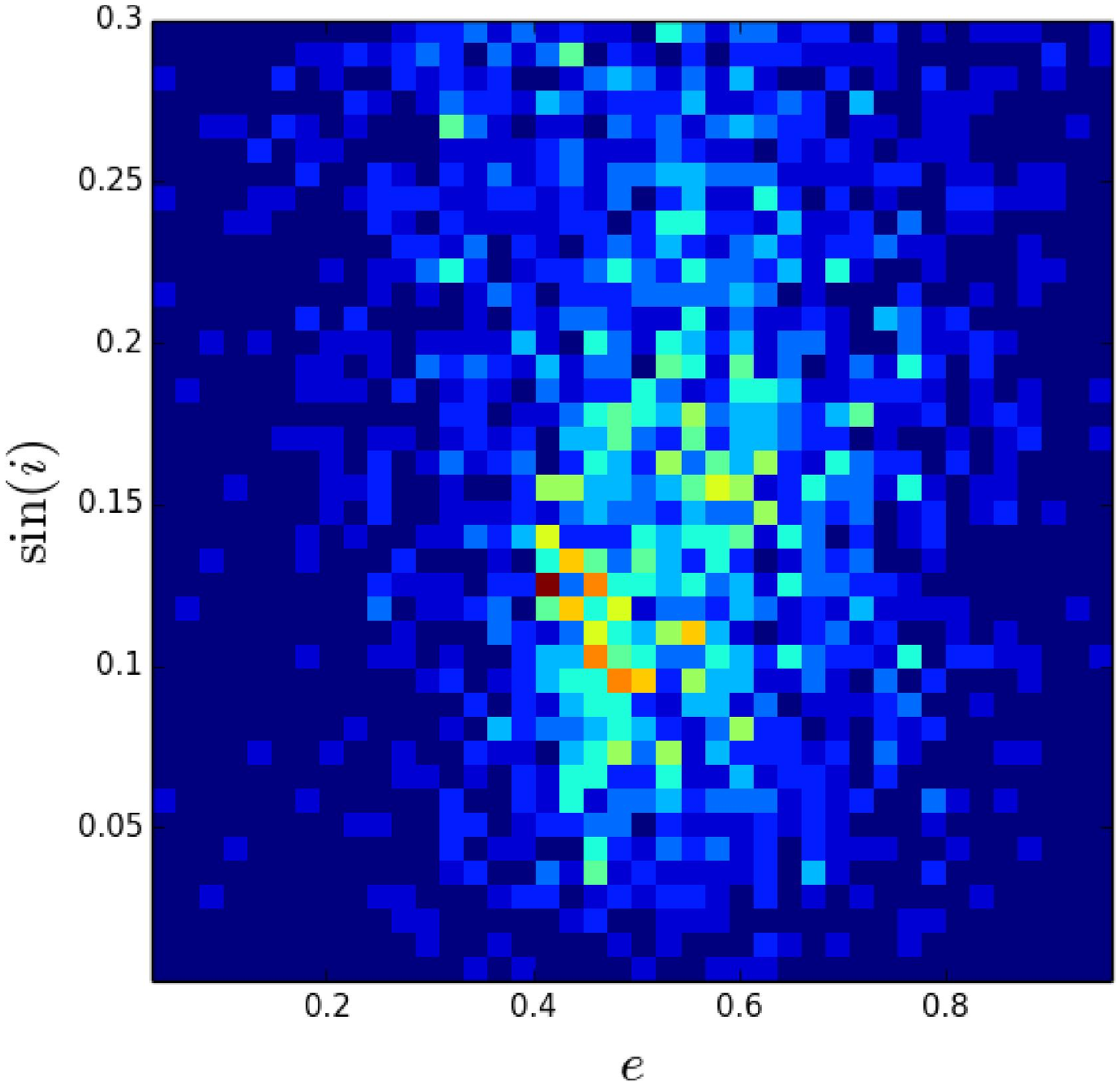}
\includegraphics[width=57mm]{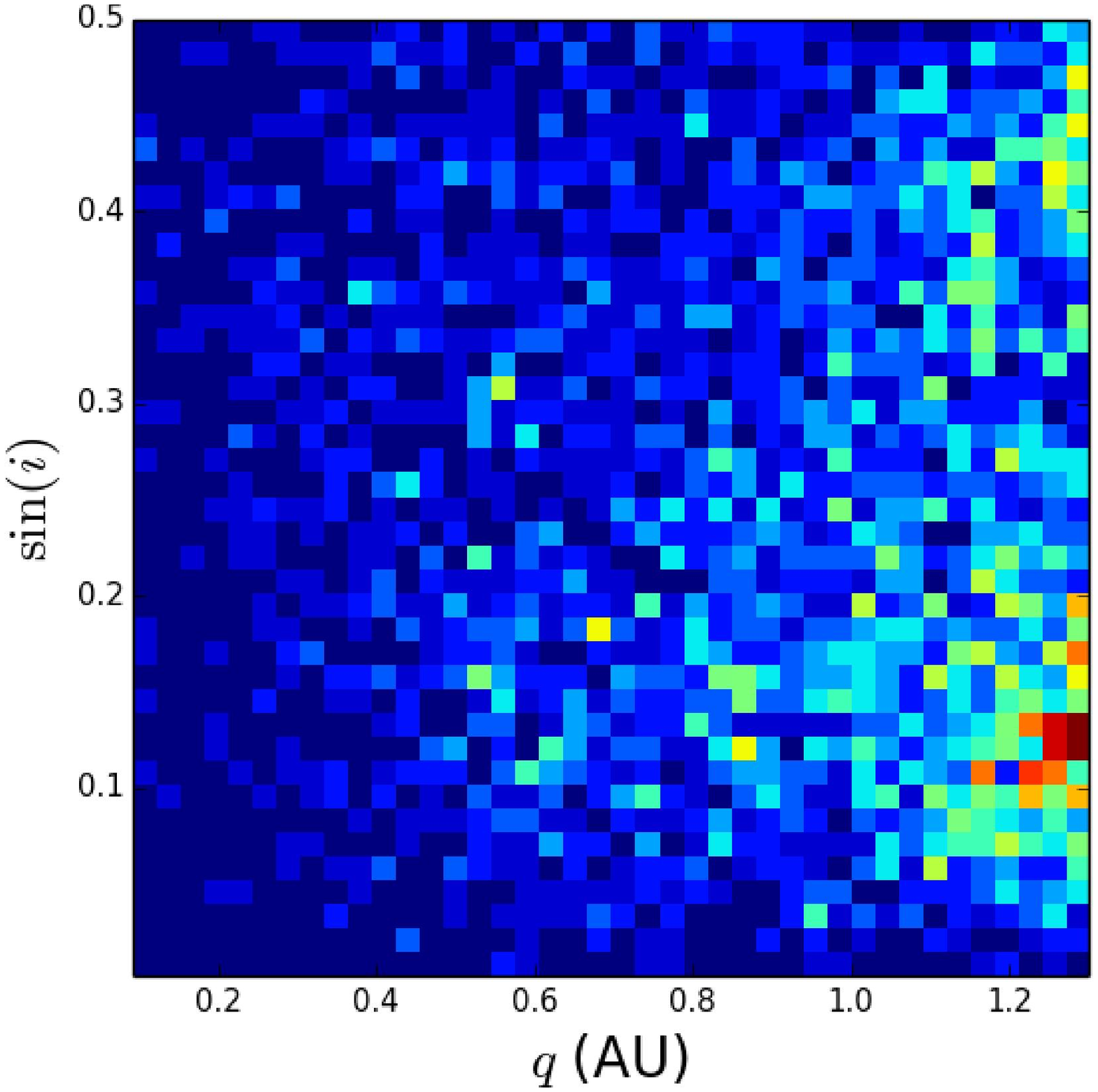}
\\\vspace{0.5cm}
\includegraphics[width=58mm]{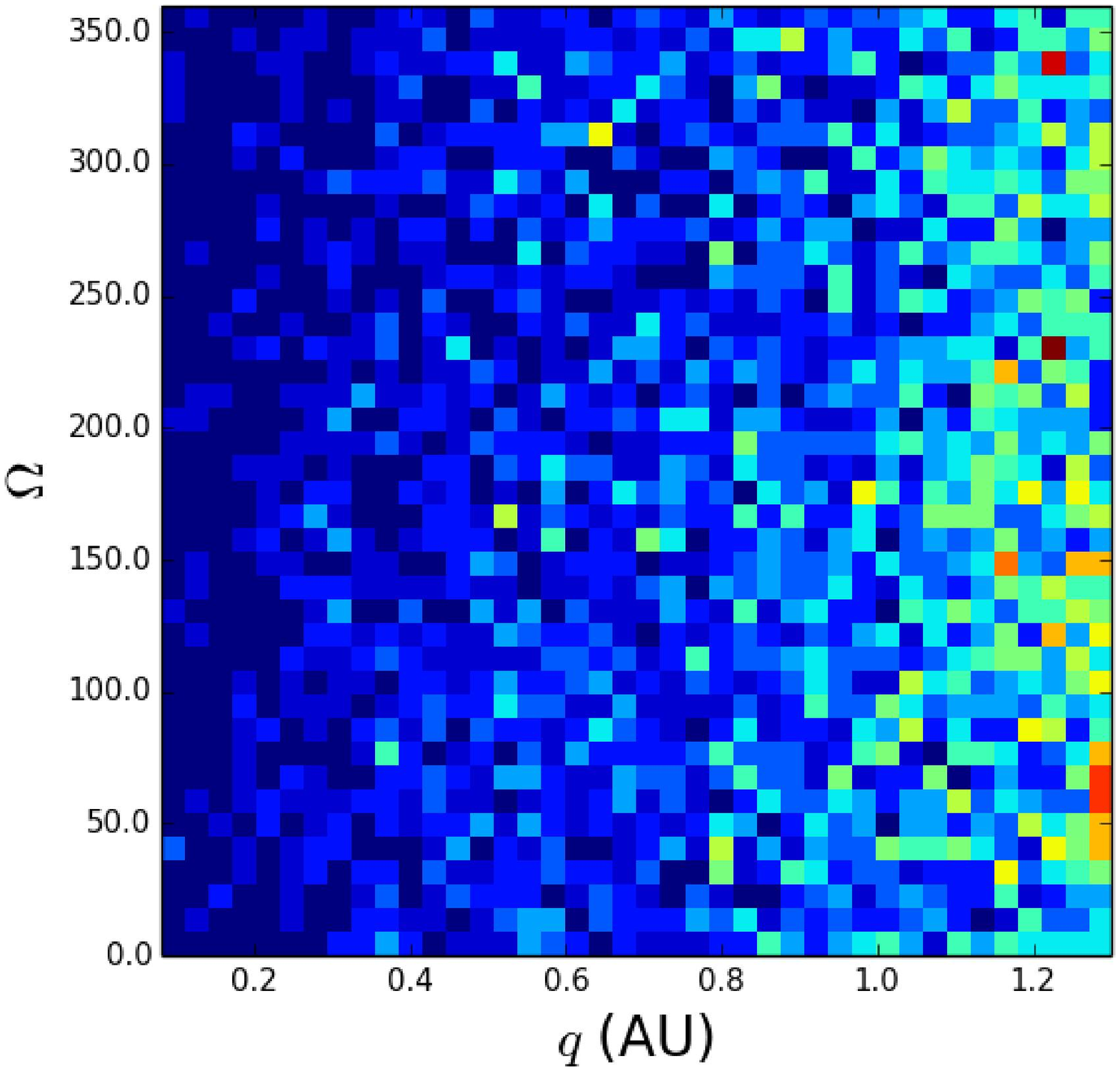}
\includegraphics[width=58mm]{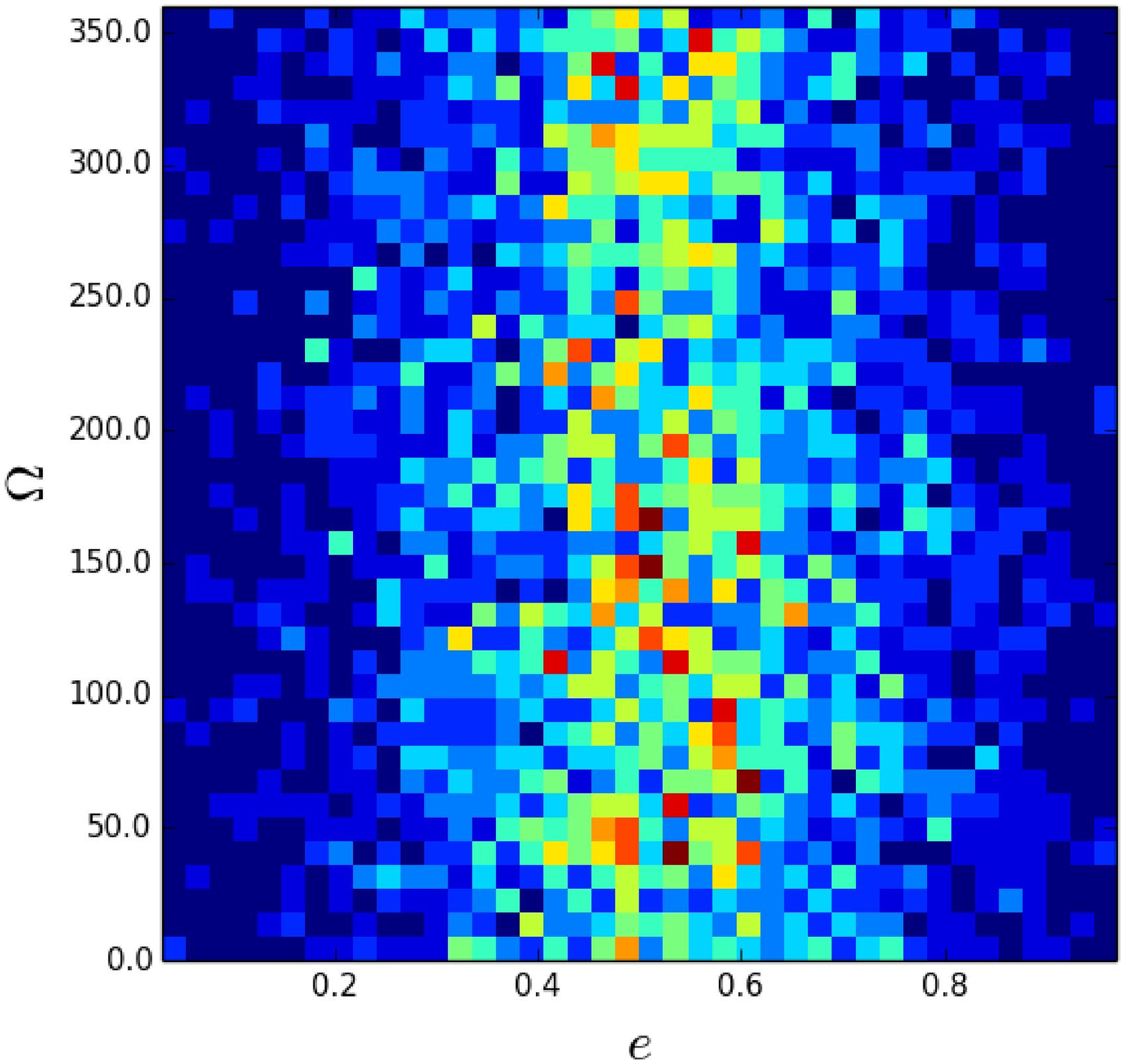}
\includegraphics[width=58mm]{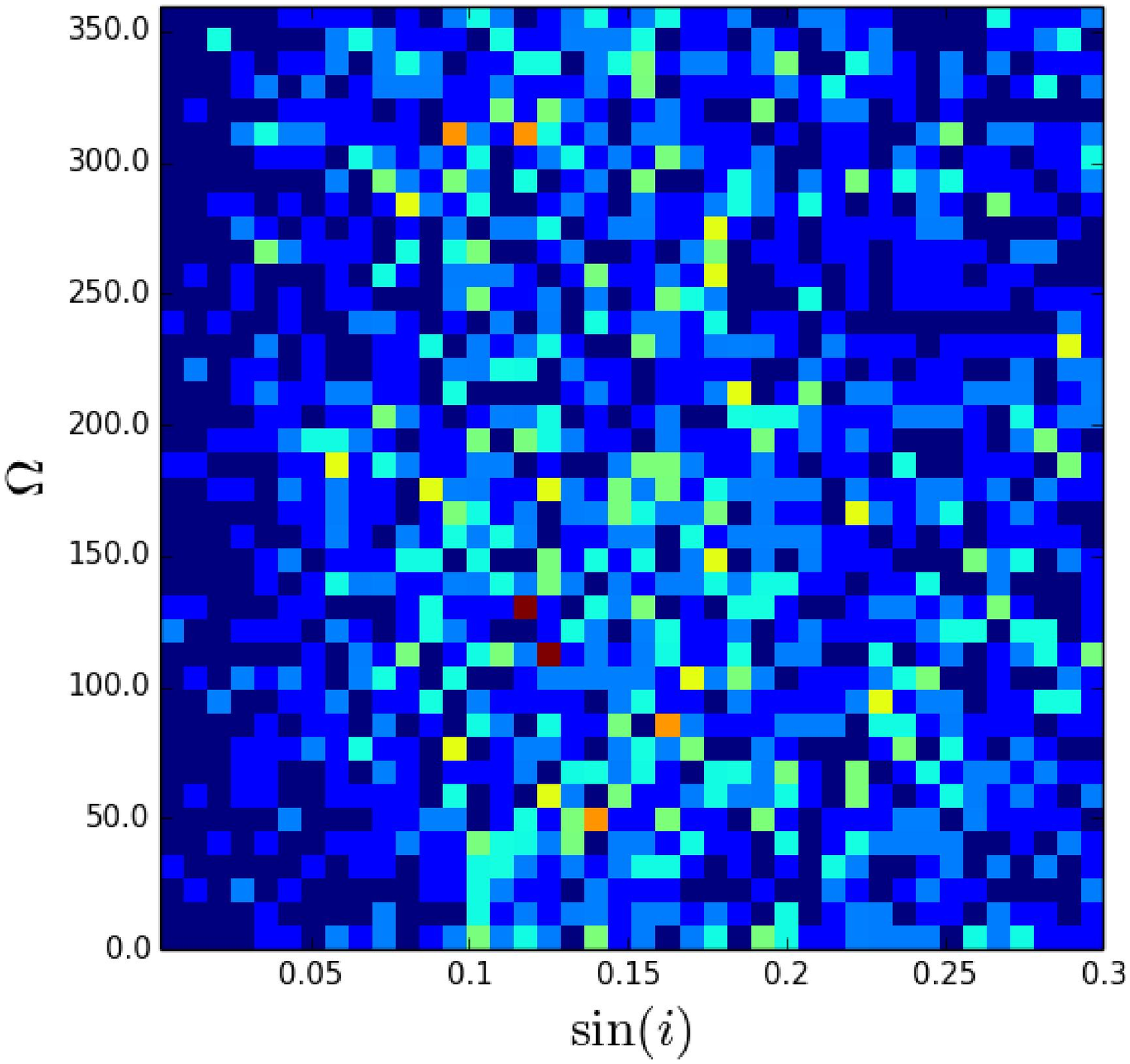}
\\\vspace{0.5cm}
\includegraphics[width=58mm]{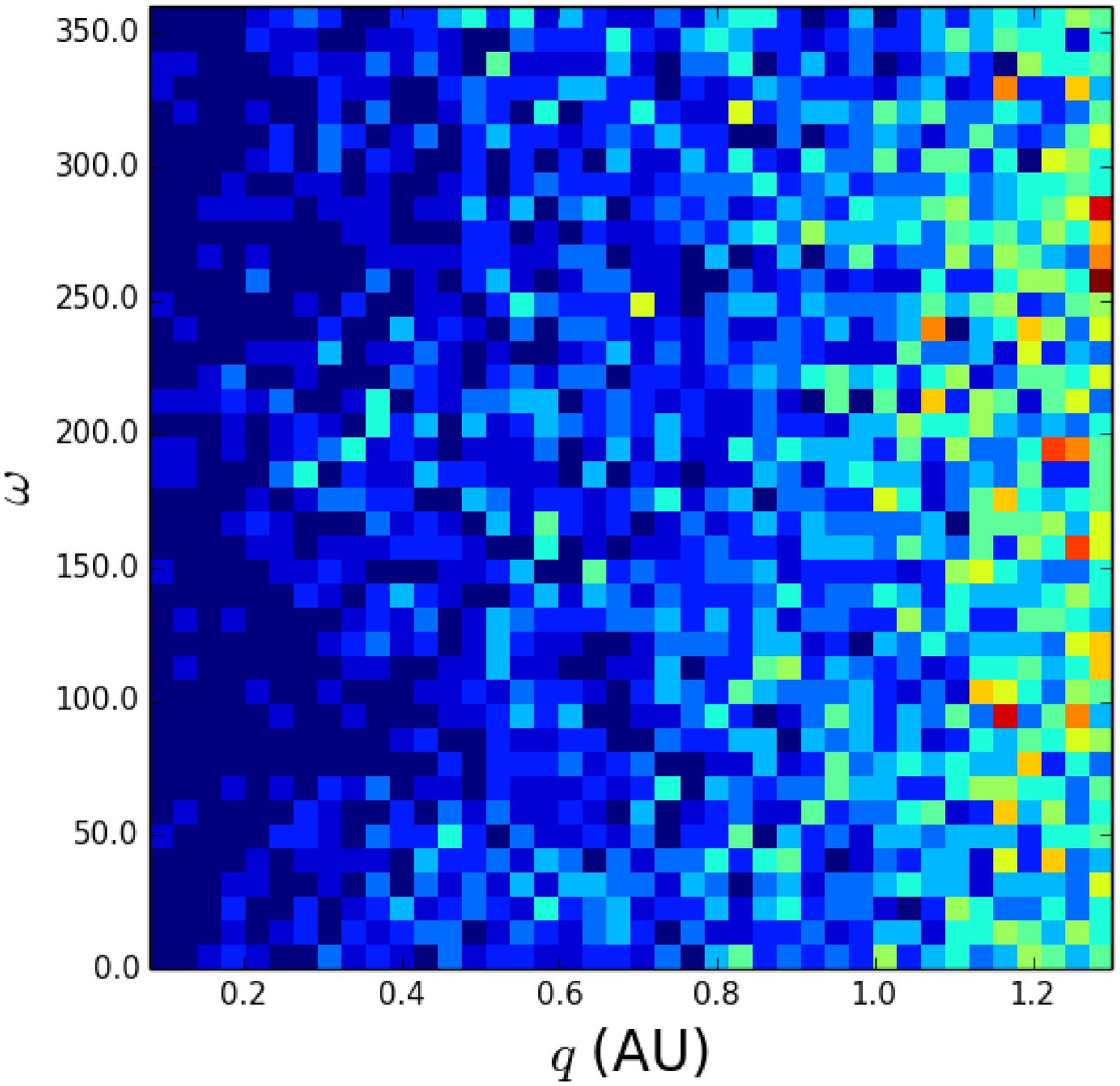}
\includegraphics[width=58mm]{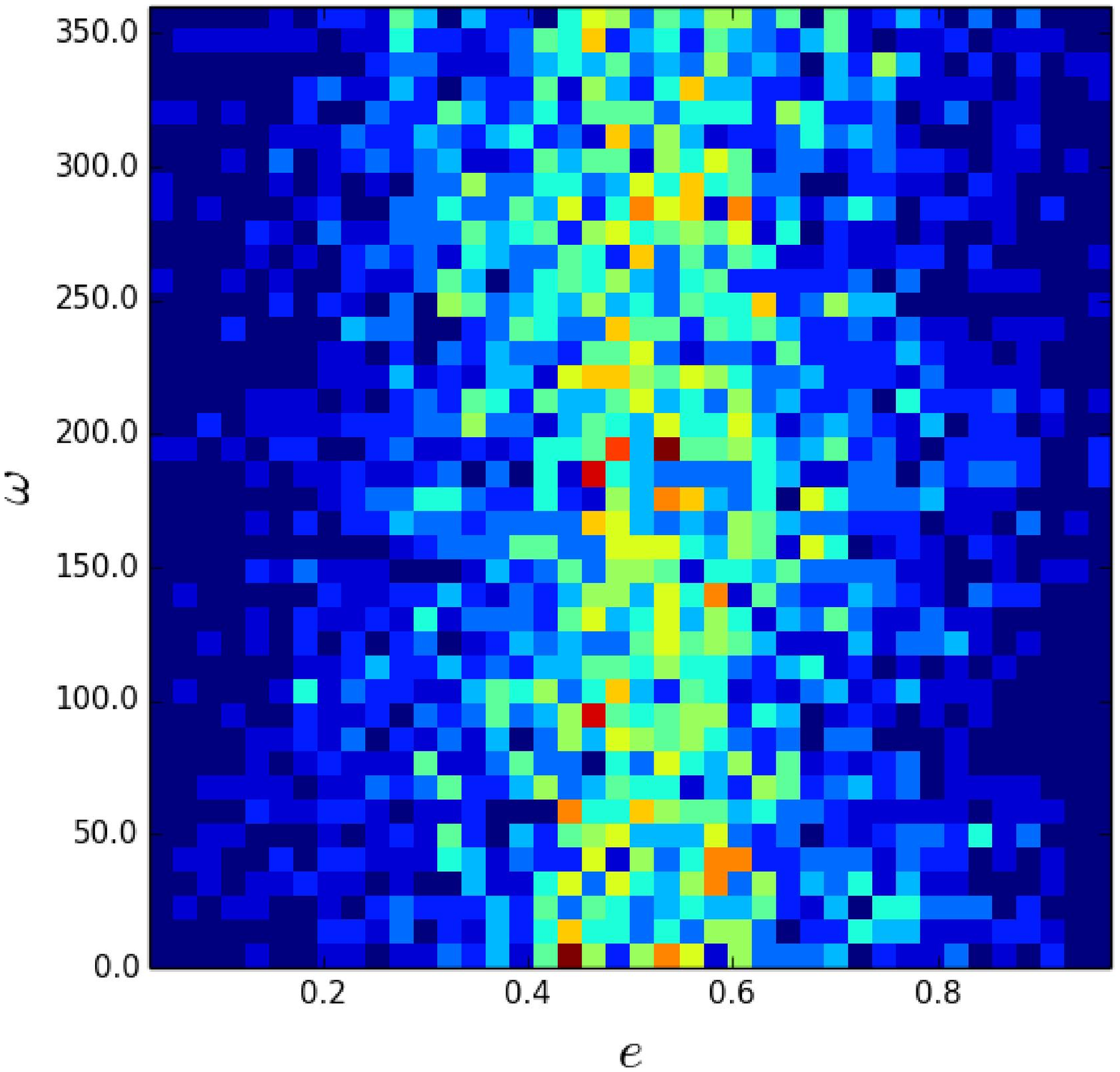}
\includegraphics[width=58mm]{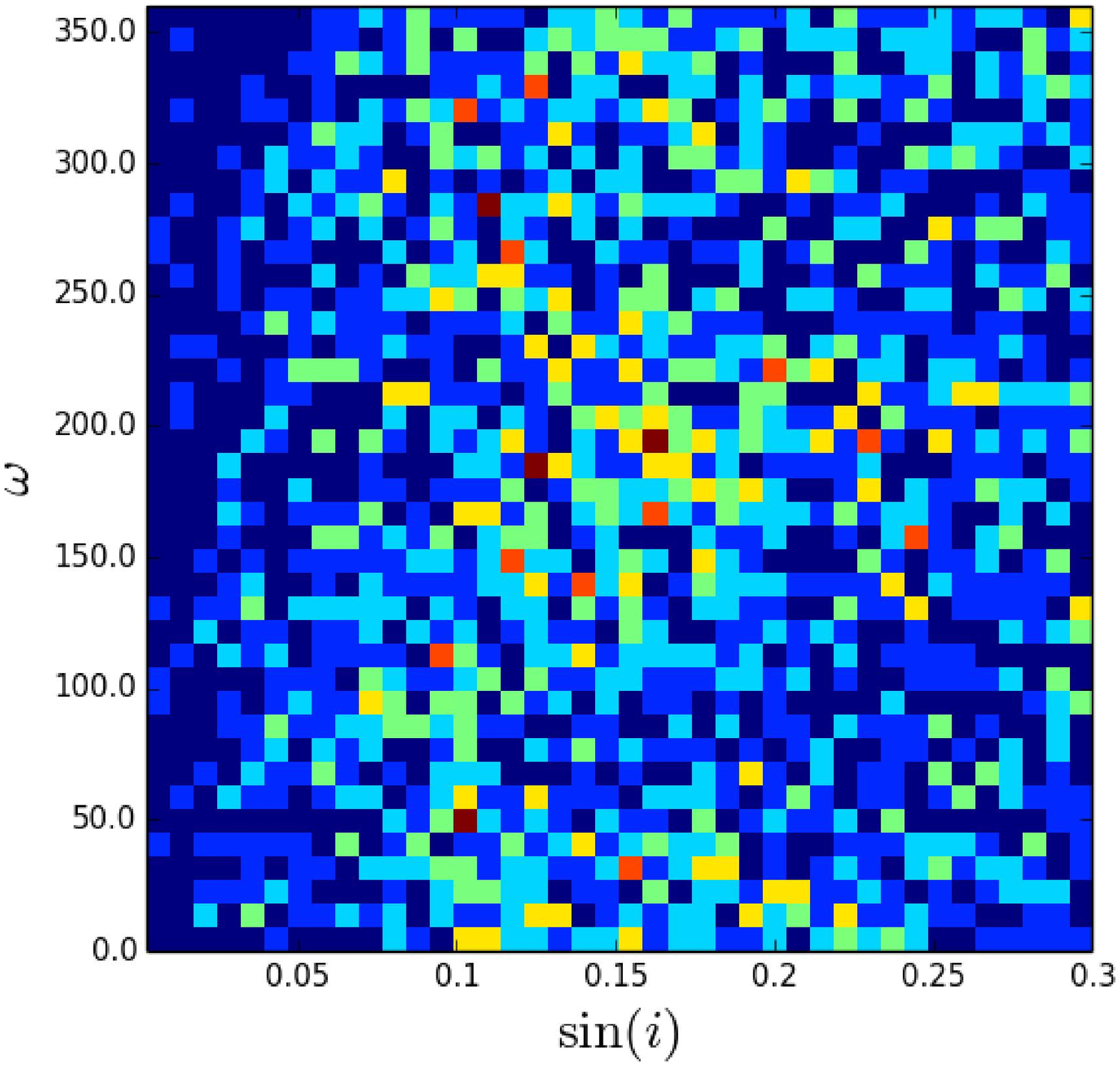}
\\\vspace{0.5cm}
   \scriptsize
  \caption{Density diagrams of the orbital properties for the (unbiased) set of 3,737 NEOs with absolute magnitude $H\lesssim 20$}.\vspace{0.2cm}
\label{fig:NEODistribution-unbiased}
\end{figure*}

In order to characterize and represent the size of the orbits, we use perihelion distance $q$ instead of semi-major axis $a$. Since NEOs are defined in terms of  $q$, ie. $q<1.3$ au, the region in configuration space enclosing these objects, has a simple rectangular geometry. Our conclusions do not depend on the selection of either $a$ or $q$ as the size parameter.

\begin{figure}  
  \centering
  \vspace{0.2cm}
  \includegraphics[scale=0.45]{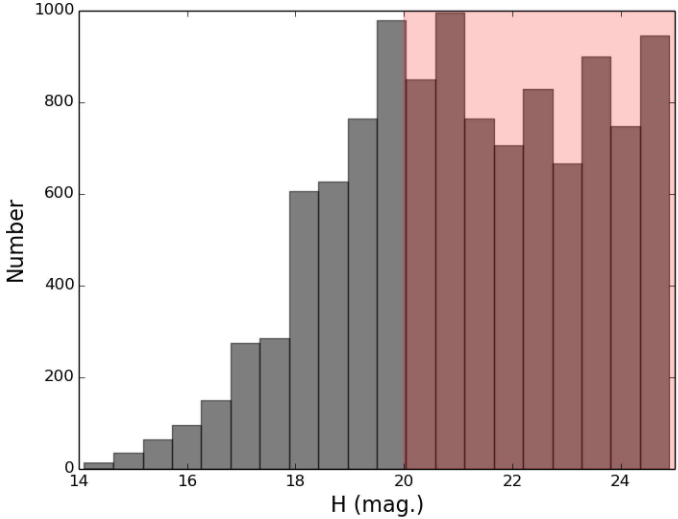}\vspace{0.5cm}
   \scriptsize
  \caption{Distribution of absolute magnitudes $H$ of NEOs.  The shaded region $H>20$ represent the values for which the sample is incomplete.}
\label{fig:HDistribution}
\end{figure}

The distribution of NEOs in \autoref{fig:NEODistribution} is affected by observational biases as described in \citet{Bottke2002}, \citet{Jeongahn2014}, \citet{Granvik2016} and \citet{Jedicke2016}. A rigorous debiasing procedure, though important for the goals of our method, is out of the scope of this paper. Still, and in order to avoid the effect of those observational biases we will use in our method NEOs having magnitudes $H\lesssim 20$ (hereafter, we will call this the ``complete set of NEOs''). In \autoref{fig:HDistribution} we show \hll{the distribution of absolute magnitude among the NEOs in the sample used in this work.  We see that that the number of NEOs grows monotonically until $H\sim 20$ where it seems to stall. Although this information alone is insufficient to \hlll{prove} that the sample is complete for $H\lesssim 20$, this limit is consistent with the results in \citet{Harris2015} and \citet{Tricarico2017} which predict more rigorously a maximum $\sim$50\% completeness for NEOs having $H\sim 20$.}

As observed from comparing densities of the samples in Figures \autoref{fig:NEODistribution} and \autoref{fig:NEODistribution-unbiased}, many of the observational biases identified by \citet{Bottke2002} and \citet{Jeongahn2014} vanish when restricting to the complete set of NEOs.  These biases include, but are not restricted to, an excess of $q=1$ and $e<0.4$ objects in the full sample as well as strong seasonal dependencies on the distribution of $\Omega$ and $\omega$.  In the unbiased (complete) sample, the  non uniformities in $\Omega$  are washed out.  However, in the case of $\omega$ (bottom row in \autoref{fig:NEODistribution-unbiased}) some slight non-uniformities remain. Both facts are in agreement with the conclusions of \citet{Jeongahn2014}.
 
\hl{Here, it is important to stress that for the simple tests of the method we present in this work, we are assuming that the spatial distribution for objects with $H<20$ (diameters larger than 0.5 km, \citealt{Jedicke2016}) is representative of the distribution of smaller objects, down to a few tens of meters.  This is not completely true as it has been recently suggested by \citet{Granvik2017}.  Both, the Yarkovsky effect and YORP cycles affect differently the orbit of small objects with respect to larger ones.  This may be reflected in significant differences in the spatial distribution of H<20 objects with respect to smaller objects.  Again, a proper consideration of all these effects is left for a future work.}

\subsection{Impact probability}
\label{subsec:probability}

The probability that a site with latitude $\phi$ and longitude $\lambda$ be impacted at $t$ is:

\begin{eqnarray}
\label{eq:ImpactProbabilityContinuous}
P\sub{imp}(\lambda,\phi;t)&=&\int_{\vesc}^{\infty}\int_{\rm 2\pi} p(\Omega,\vimp;t)\,d\Omega\,d\vimp
\end{eqnarray}
 
Here $p(\Omega,\vimp;t)\,d\Omega\,d\vimp$ is the probability that at time t,  an object comes from a direction $\Omega:(A,z)$, inside a solid angle $d\Omega$ and with incoming speed between $\vimp$ and $\vimp+d\vimp$. Here we should not confuse the solid angle $\Omega$ with the longitude of the ascending node. GRT is essentially a Monte Carlo method designed to evaluate this integral.  

In the discrete limit \autoref{eq:ImpactProbabilityContinuous} is written as:

\begin{eqnarray}
\label{eq:ImpactProbability}
P\sub{imp}(\lambda,\phi;t)&\propto& \sum_j P\sub{in}(A_j,z_j,v\sub{imp,j};t)
\end{eqnarray}

\noindent where the sum is over all the test particles used to evaluate the integral. $A_j,z_j,v\sub{imp,j}$ are the impact conditions (initial conditions for the backward-integration) of the $j$th test particle and $P\sub{in}(A_j,z_j,v\sub{imp,j};t)$ is the probability of having those initial conditions. 

Backward-integration maps each impact direction and speed $(A,z,\vimp)$ at a given geographical site and time $t$, into one point in configuration space with osculating elements $q(A,z,\vimp,\lambda,\phi;t)$, $e(A,z,\vimp,\lambda,\phi;t)$, $i(A,z,\vimp,\lambda,\phi;t)$, $\Omega(A,z,\vimp,\lambda,\phi;t)$,  $\omega(A,z,\vimp,\lambda,\phi;t)$.  We will call the orbit defined by these elements, the ``Asymptotic orbit'' of the test particle (see \autoref{fig:GRT}). 

The key assumption of the method is that the probability of having initial conditions $(A,z,\vimp)$ at a given time and location is proportional to:

\begin{eqnarray}
\label{eq:ImpactProbabilityDiscrete}
P\sub{in}(A,z,\vimp;t)&\propto& R(q,e,i,\Omega,\omega).
\end{eqnarray}

where $R(q,e,i,\Omega,\omega)$ is the number density of parent bodies in configuration space as measured at the orbital elements of the asymptotic orbit.

\subsection{Flux correction to impact probability}
\label{subsec:flux}

 {
The Earth is a moving target. This fact is responsible for a kinematic ``focusing'' or ``defocusing'' effect that alters the flux of objects coming from the stationary distribution $R(q,e,i,\Omega,\omega)$ as calculated in the preceding section.  

If the impactors were at rest in interplanetary space, the flux of particles coming from the apex (the direction of Earth movement) will be larger than in other directions (focusing effect). In general the flux would only depend on the angle $\theta\sub{apex}$ between the particle radiant and the apex direction.

But NEOs are not at rest. Their proper motion in the vicinity of the Earth depends in a complicated way on their distribution $R(q,e,i,\Omega,\omega)$.  Still, boundary conditions for the flux at three extreme cases, namely $\theta\sub{apex}=0\deg, 90\deg$ and $180\deg$, can be derived.

In order to have an impactor with an oncoming $\theta\sub{apex}\sim 0\deg$ or $\theta\sub{apex}\sim180\deg$, its heliocentric orbit should fulfill one of two conditions: (1) to be nearly circular ($e\approx 0$, $q\approx 1$ \hll{or $Q\approx 1$}), or (2) to have a perihelion argument close to 0 or 180$\deg$ \hll{(while having $q\approx 1$ or $Q\approx 1$ respectively).  In both cases, and to have an impact specifically close to apex and antapex, the relative inclination must also be very low, ie. $i\approx 0$}.  The number of NEOs fulfilling the first condition is very low ($\lesssim 1\%$ of NEOs have eccentricities $e<0.1$).  On the other hand, since the distribution of perihelion arguments is very flat (though not uniform), the probability of having a particular value of $\omega$ (0$\deg$ or 180$\deg$) is also very low (eg. $\sim 1\%$ for $\omega<5\deg$).

In summary the apex  $\theta\sub{apex}=0\deg$ and antapex $\theta\sub{apex}=180\deg$ directions in the sky at any location on Earth, will see the lowest fluxes while the perpendicular directions, $\theta\sub{apex}\sim 90\deg$, will see the largest number of incoming particles. We call this a ``defocusing'' effect.

In other directions, the flux of incoming particles will depend in a complex way on $R(q,e,i,\Omega,\omega)$ and the properties of the Earth's orbit.  Moreover, the observation of sporadic meteors suggests that this flux could depend in complicated ways, not only on $\theta\sub{apex}$, but also on the longitudinal angle $\lambda\sub{apex}$ around the apex-antapex direction \citep{Campbell2008}.

In general the resulting initial condition probability in Eq. \ref{eq:ImpactProbabilityDiscrete}) should be corrected by these kinematic effects:

\begin{eqnarray}
\label{eq:ImpactProbabilityDiscreteFlux}
P^{*}(A,z,\vimp;t)&\propto& f(\theta\sub{apex},\lambda\sub{apex}) R(q,e,i,\Omega,\omega).
\end{eqnarray}

Modeling the kinematic effects is not trivial.  In this initial approach we will use the simplest ansatz. In \autoref{fig:DirectionsDistribution} we present the observed distribution of $\theta\sub{apex}$ as obtained from observations of large fireballs and bolides reported by NASA Near Earth Objects program\footnote{\url{http://neo.jpl.nasa.gov/fireballs}}. After correcting the distribution for the area effect, the shape of the flux can be modeled with the piecewise trigonometric function:

\begin{eqnarray}
f(u=\cos[90-\theta\sub{apex}];a,b)&=&
\left\{
\begin{array}{ll}
u^a & -1\leq u < 0; \\
u^b & 0\leq u \leq 1.
\end{array}
\right., 
\label{eq:Flux}
\end{eqnarray}

with $a$ and $b$ free parameters. The observed shape of the flux in \autoref{fig:DirectionsDistribution} is reproduced using $a=1$ and $b=0.5$. 

\begin{figure}  
  \centering
  \vspace{0.2cm}
   \includegraphics[scale=0.45]{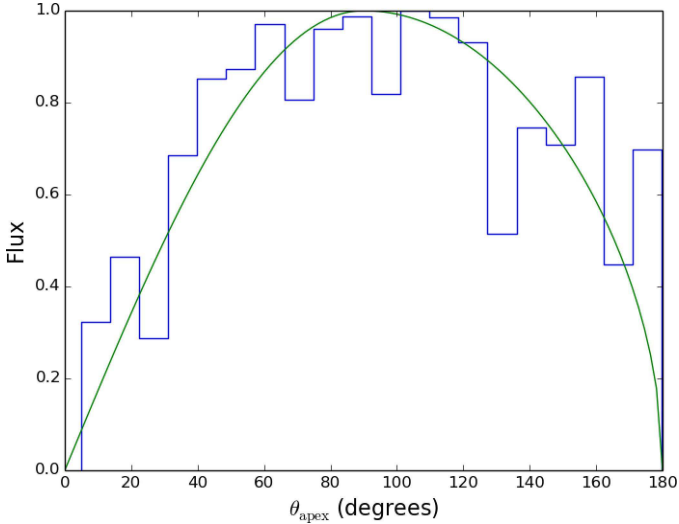}\\\vspace{0.5cm}
  \scriptsize
  \caption{Distribution of incoming directions with respect to apex (histogram) for the objects in the NASA's fireball and bolide database.  The distribution can be model with a piecewise trigonometric function (solid line).}
\label{fig:DirectionsDistribution}
\end{figure}

In the following we will assume a flux correction to impact probability that follows this simple distribution and that does not depend on $\lambda\sub{apex}$.  A more complex model is left for future investigations.

\subsection{Density of bodies in configuration space}
\label{subsec:density}

Calculating the density $R(q,e,i,\Omega,\omega)$ from a discrete sample of several thousands of orbital elements that do not span the entire configuration space is challenging. Different numerical techniques have been devised in areas with similar challenges, ranging from cosmology to hydrodynamics (for a recent review see eg. \citealt{Price2012}). For GRT we implement a formulation widely used and tested in Smooth Particle Hydrodynamics \citep{Price2012}.  

The number density objects around a point $\vec{x}\equiv(q,e,i,\Omega,\omega)$ in configuration space is given by:
\begin{eqnarray}
R(\vec x) &=& \sum_k W(||\vec{x}-\vec{x}_k||,h),
\label{eq:SmoothingFunction}
\end{eqnarray}
 
where $||\vec{x}-\vec{x}_k||$ is the generalized ``distance'' between points in configuration space (see below) and $h$ is a scale parameter.  $W(||\vec{x}-\vec{x}_k||,h)$ is called the {\it smoothing kernel}, and represents the function that allows a soft transition from the discrete to the continuous regime. 

Detailed numerical experiments have demonstrated that the best density estimates are obtained using the so-called {\it B-spline kernel} \citep{Price2012}:

\begin{eqnarray}
W(||\vec{x}-\vec{x}_k||,h)&\equiv& {\frac{1}{h^3}} w(u=||\vec{x}-\vec{x}_k||/h)  \\ &=&\frac{\sigma}{4h^3}
\left\{
\begin{array}{ll}
(2-u)^3-(1-u)^3, & 0\leq u < 1; \\
(2-u)^3, & 1\leq u < 2; \\
0, & 1\leq u \geq 2,\nonumber
\end{array}
\right.
\end{eqnarray}
 
where $\sigma$ is a normalization constant.

Distances in configuration space $||\vec{x}-\vec{x}_k||$ are computed using the \citet{Zappala1990} metric (hereafter Z-metric). For this purpose we use the  parametrization introduced in \citet{Rozek2011}:

\hll{
\beq
\label{eq:Z-metric}
\begin{array}{lll}
(D_Z/n_m a_m)^2 & = &
\frac{5}{4} (a-a_m)^2/a_m^2 + 
2 (e - e_k)^2 + \\ 
&  & + 2 (\sin i - \sin i_k)^2 +\\
&  & + 
10^{-4} (\Omega - \Omega_k)^2 +
10^{-4} (\varpi - \varpi_k)^2
\end{array} 
\eeq
}

Here $a_m=(a+a_k)/2$ is the average semi-major axis, $n_m$ is the corresponding orbital mean motion \hll{and $\varpi=\Omega+\omega$ is the longitude of the perihelion}.

 The Z-metric is particularly well suited for our purposes  {since it is easy to implement numerically and includes the orbital elements $\Omega$ and $\omega$.}  Moreover, the metric has been successfully used for comparing orbital elements of asteroids ($e<1$) and to perform cluster analysis in configuration space, which is similar in nature to the calculations required here. We have tested another metric, ie. the Drummond metric, and found no significant differences in the resulting probability distributions, modulo normalization constants.

\section{Results}
\label{sec:results}

In the direction of illustrating the application of the GRT method and compare its results against observations, we have devised 4 numerical experiments:

\begin{enumerate}
\item To compute the relative impact probability for a discrete set of geographical locations at the same date and time.
\item To compute the relative impact probability for a discrete set of geographical locations at different dates and times.
\item To create a map of relative impact probability at specific dates and times.
\item To compute other statistical properties, such as the distribution of impact speeds from the probabilities computed at a given location, date and time.
\end{enumerate}

Each experiment will illustrate specific aspects of the method and will clarify several open issues from preceding sections.

\subsection{Discrete locations at the same date and time}
\label{subsec:discrete_locations}

Our first experiment is the computation of relative impact probabilities at the date and time of one of the events that inspired us to develop GRT, namely the Chelyabinsk impact.  The event occurred on February the 15th, 03:20:34 UTC.  

We will calculate the relative probability of 3 specific locations: Chel\-yab\-insk, Russia (lon. 63.5$\deg$, lat.  54.4$\deg$), Honolulu, Hawaii (lon. -157.8$\deg$, lat. 21.3$\deg$) and Antananarivo, Madagascar (lon. 47.5$\deg$, lat. -18.9$\deg$). Madagascar and Hawaii locations are selected for this experiment since at the time of the event they were, respectively, close to the projection on the Earth' surface of the apex and the anteapex. Hereafter we will call these locations the ``geographic apex'' and ``geographic antapex'' as opposed to the celestial apex and antapex.  The colatitude with respect to these direction we will call $\theta\sub{apex}\sup{geo}$ as opposed to the already defined angle $\theta\sub{apex}$ that is measured in the sky.

To perform the GRT analysis we must first generate random impact incoming directions and impact speeds using the methods described in \autoref{subsec:initial}. For our experiments we generate, using Poisson sampling, random directions with a minimum angular separation of 20$\deg$ in the sky. This gives us 27 possible pairs of azimuth and zenith angles (in Poisson sampling the number of random points is a product of the algorithm and cannot be set a priori). A lower minimum angular separation produces many more sampling points increasing the computational cost; a larger minimum will reduce considerably the resolution of the method. Additionally we use 30 equally spaced impact velocities in the interval [11.1,\hl{43.6}] km/s.  Together, a total of 810  ($27\times 30$) different initial conditions for our test particles were used at each geographic location. \hl{We have tested the method with a much larger resolution (and a correspondingly larger computational cost), namely, using 110 different directions (separation of 12$\deg$ in the sky) and 52 initial speeds, without noticing a significant improvement in the relative probabilities.}

\autoref{fig:PointMaps} shows the asymptotic orbital elements of the test particles thrown from the three test sites. We call this type of diagrams the ``GRT fingerprint'' of a test location.

\begin{figure*}  
  \centering
  \vspace{0.2cm}
   \includegraphics[width=180mm]{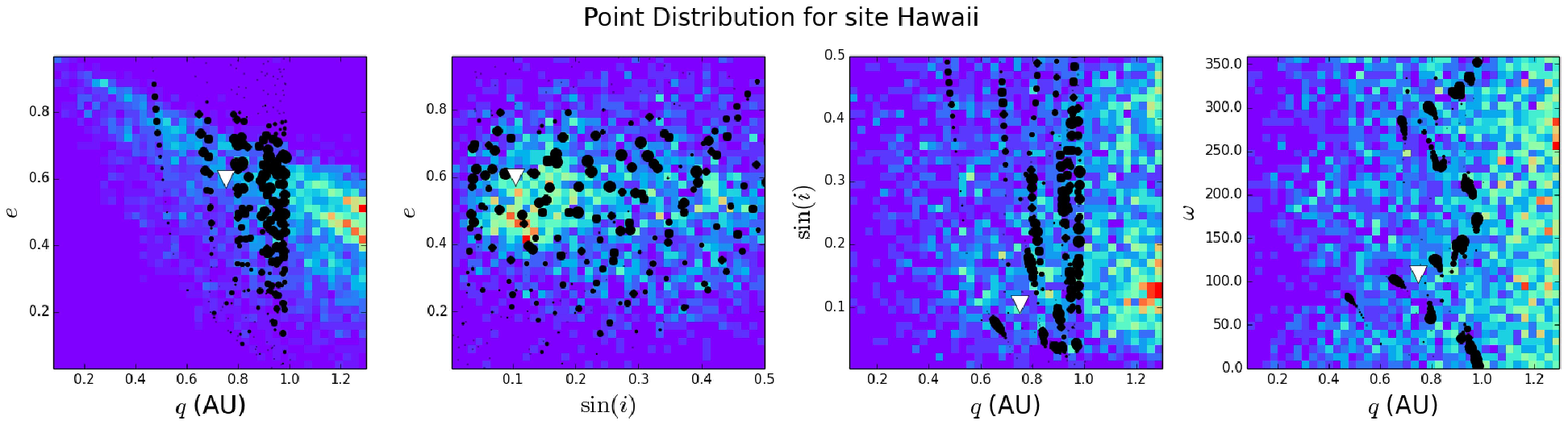}\\\vspace{0.3cm}
   \includegraphics[width=180mm]{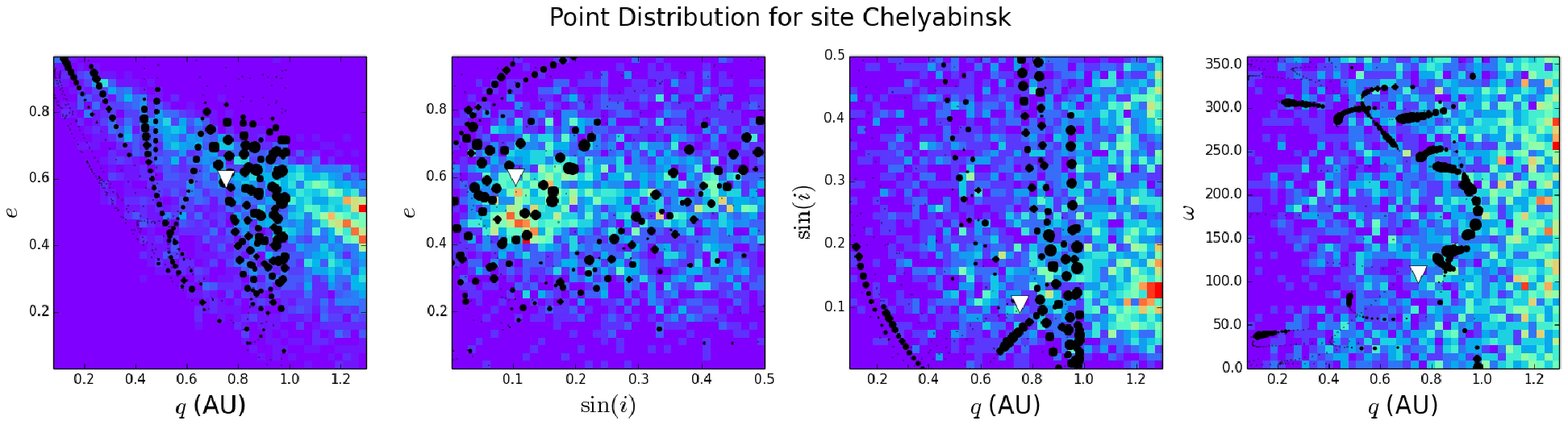}\\\vspace{0.3cm}
   \includegraphics[width=180mm]{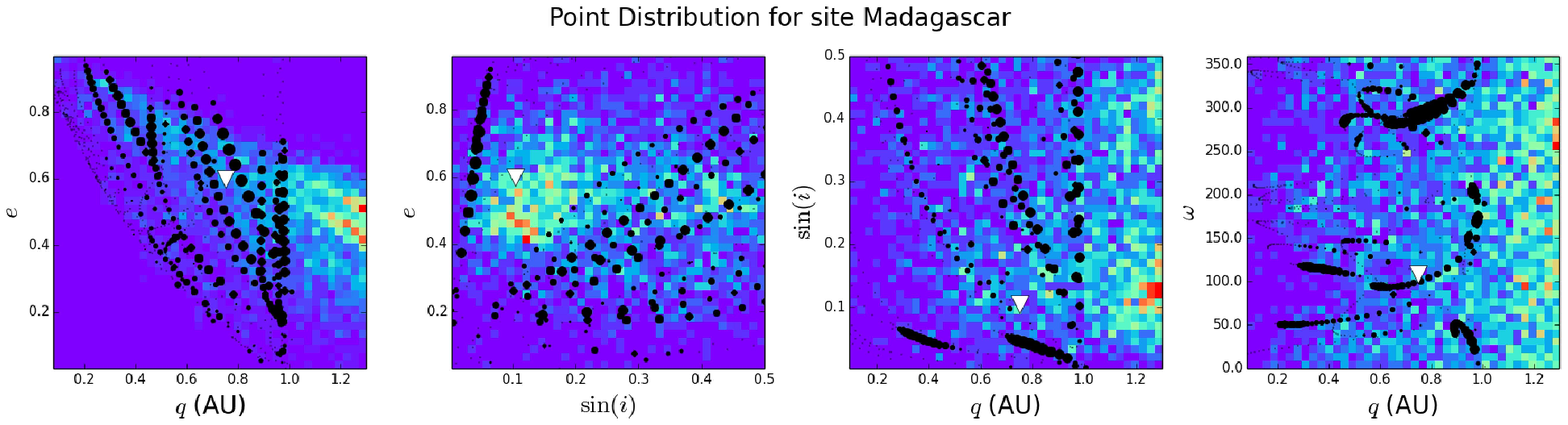}\\\vspace{0.3cm}
   \scriptsize
  \caption{Asymptotic orbital elements of test particles (black circles) thrown from 3 different sites at the date and time of the Chelyabinsk impact (GRT fingerprints).  \hll{The inverted white triangle indicates the orbital elements of the actual Chelyabinsk impactor}.  {The size of each black circle is proportional to the probability $P^{*}\sub{ini}$ of the initial condition for test particle}. Sites are sorted from top to bottom, in descending order of impact probability. Color in this diagram has the same meaning as in the density plots in \autoref{fig:NEODistribution}.\vspace{0.2cm}}
\label{fig:PointMaps}
\end{figure*}

The GRT probability relies on 3 free parameters: the scale parameter of the smoothing function $h$ (\autoref{eq:SmoothingFunction}) and the two exponents $a$ and $b$ of the flux correcting function (\autoref{eq:Flux}).  We find that $h=0.2$, which is close to the average Z-distance (\autoref{eq:Z-metric}) between the NEOs in the unbiased sample, works well for estimating accurately the density while keeping the computational cost low. For the flux correcting function we use in this experiment the values $a=1$ and $b=0.5$ that reproduces the fireball $\theta\sub{apex}$ distribution (\autoref{fig:DirectionsDistribution}).

Computing the normalization of Probability $P\sub{imp}$ in \autoref{eq:ImpactProbabilityDiscrete} is challenging.  Instead, we can calculate the relative probability, ie. the ratio of impact probability at the site of interest and at a reference site, both at the the same date and time:

$$
p\sub{site}(t)=\frac{P\sub{imp,Site}(t)}{P\sub{imp,Ref.Site}(t)}
$$

For reasons that will be clear below, we choose the geographic antapex as the reference location.

Under these conditions the relative probability for our three sites are: $p\sub{Hawaii}=0.96$, $p\sub{Chelyabinsk}=0.60$ and $p\sub{Madagascar}=0.39$.  

Probabilities match well the GRT fingerprint of each location. In the case of Hawaii (the closest site to  {the geographic} antapex), the GRT fingerprint  occupies high density regions in parameter space (see the top row in \autoref{fig:PointMaps}). On the other hand, test particles thrown from Madagascar (which is close to the geographic apex), have asymptotic orbits with perihelion distances systematically lower than 1 au that are less frequent among NEOs.

Interestingly, although Chelyabinsk has a relative probability 50$\%$ larger than Madagascar, both sites have similar GRT fingerprints. This is an effect of the flux correction. In Chelyabinsk, whose apex geographic colatitude is $\theta\sub{apex}\sup{geo}=75.5\deg$ many more particles would come from directions $\theta\sub{apex}\sim 90\deg$ that have a largest flux. As a consequence, two test particles in Chelyabinsk and Madagascar having similar asymptotic orbits could weight differently when computing the contribution to impact probability at each site.

\subsection{Discrete locations at different dates and times}
\label{subsec:discrete_times}

Strictly speaking, impact probability $P\sub{imp}(\lambda,\phi;t)$ is a three-variate function.  We cannot attempt to test the distribution using a single date and time (only one event happens at a time). Neither should we attempt to test the distribution by studying the geographical distribution of impacts observed in an extended period of time.

In order to statistically test if a set of impact events follow $P\sub{imp}$, we must compute the relative probability of every event at the location and time when they occurred.

We have performed this calculation for the impact sites of 394 fireballs in the  NASA bolide database. In \autoref{fig:FireballPDistribution} we show the distribution of relative impact probabilities $p\sub{site}(t)$ as a function of $a$ and $b$. 

We would expect that if impacts follow the probability computed with GRT, most of the events would have large relative probabilities. This is exactly what happens when the $b$ exponent is increased.

The relative probability of fireballs increases with $b$ due to \hl{a (non-trivial) combination of three} effects: (1) an intrinsic increase in the relative probability of the fireballs locations and dates, (2) a decrease in the probability of the reference site, i.e. the geographical antapex and \hl{(3) a flatter probability across most geographical sites}.  The first \hl{and third effect} are probably responsible for the decrease in dispersion of the probability distribution when using $b=3$ with respect to lower values of $b$.  This is the desired effect when thinking about the optimal set of parameters ``fitting'' the actual spatio-temporal distribution of impacts.  The second effect is responsible for \hl{the} shift in the peak position of the distribution\hl{, which seems to be only a numerical artifact}. \hl{The effect that the third effect has on relative probabilities, namely, the fact that the probability distribution be flatter due to larger values of $a$ and $b$, should be investigated in more detail and is left for a future paper.}

Although this is not a rigorous fitting procedure, we will assume hereafter that $a=3$ and $b=3$ are the best suited values of the flux correcting function that describe the spatio-temporal distribution of impact probability.

\begin{figure}  
  \centering
  \vspace{0.2cm}
   \includegraphics[scale=0.45]{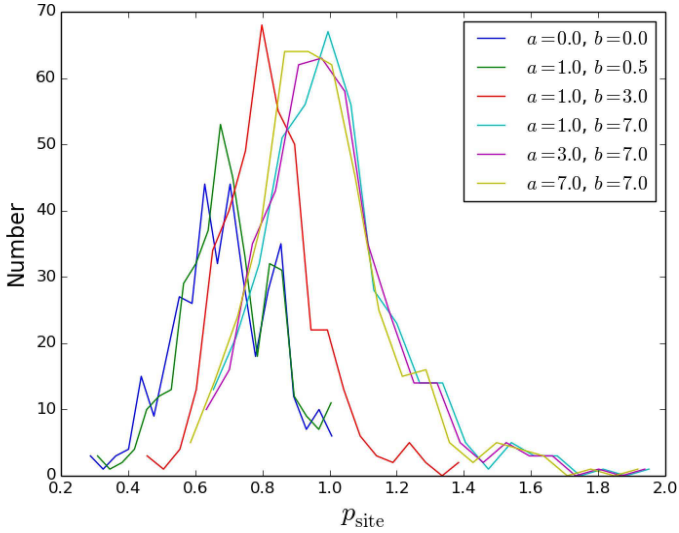}\\\vspace{0.5cm}
  \scriptsize
  \caption{Distribution of relative probability for sites and times of 394 fireballs in the NASA bolide database.  Different curves correspond to different sets of the flux correcting function parameters.\vspace{0.2cm}}
\label{fig:FireballPDistribution}
\end{figure}

\subsection{Probability maps}
\label{subsec:maps}

With a fiducial set of values for the parameters of the impact probability, we can attempt now to compute the instantaneous geographical distribution of impacts at the date and time for several key impact events.

For that purpose we repeat the procedure in \autoref{subsec:discrete_locations}, but now for 491 random Poisson-generated locations (minimum separation of 7$\deg$ or 780 km). 

Relative probability maps at the date and time of the Chelyabinsk event are shown in \autoref{fig:ProbabilityMapChelyabinsk}.

\begin{figure*}
  \centering
  \vspace{0.2cm}
   \includegraphics[width=115mm]{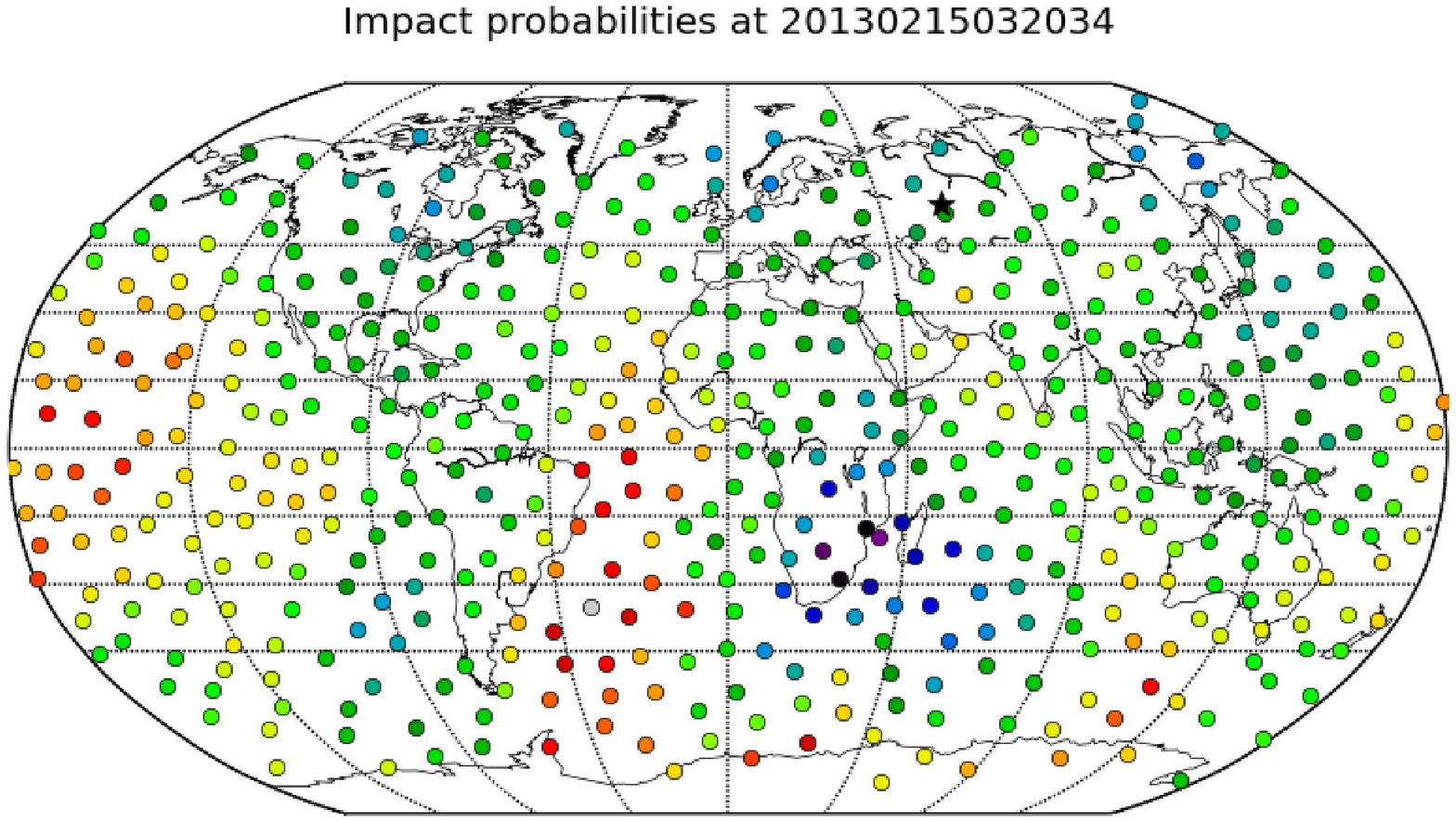}\\\vspace{0.2cm}
   \includegraphics[width=115mm]{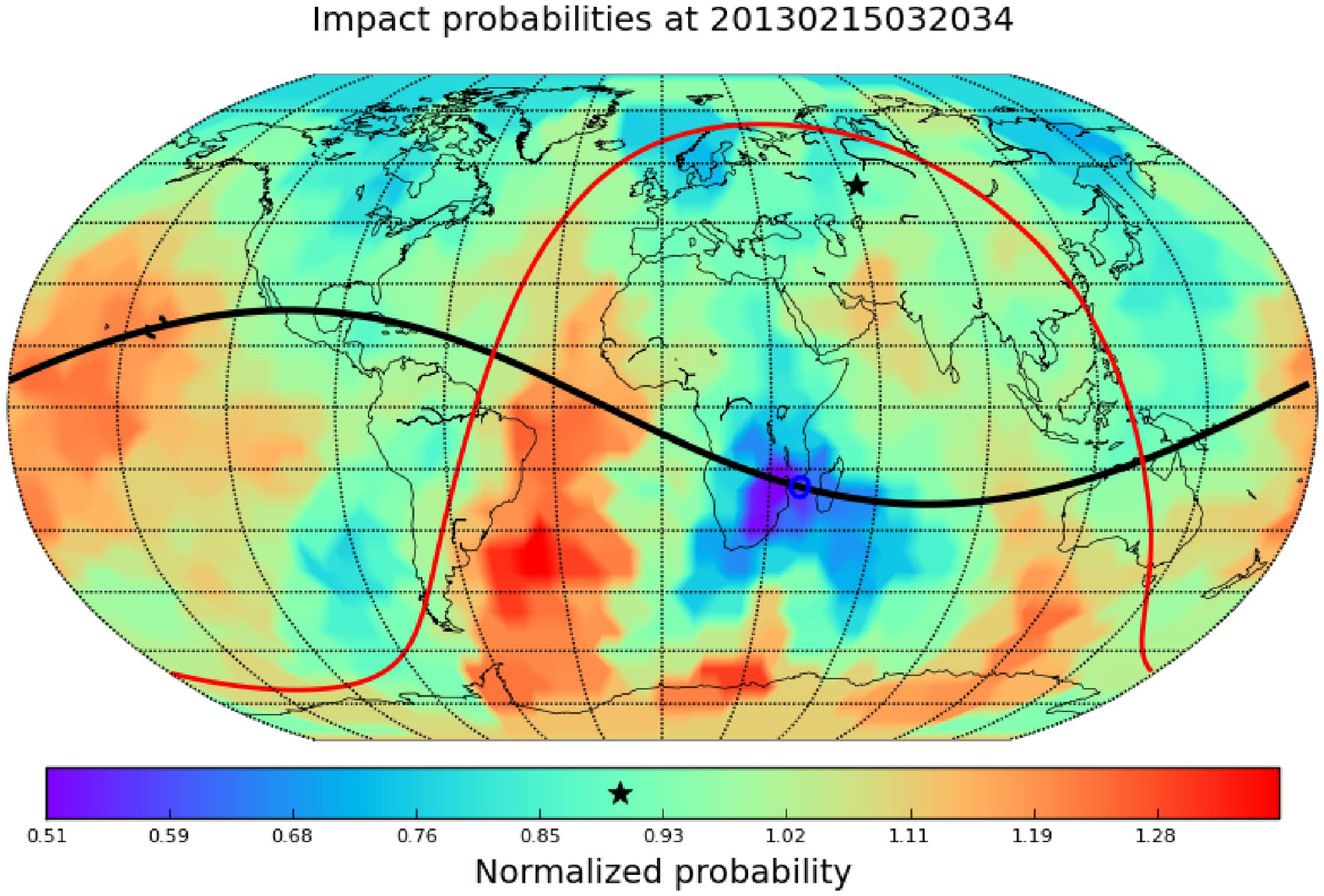}\\\vspace{0.2cm}
   \scriptsize
  \caption{Map of relative impact probability across the Earth's surface at the date and time of the Chelyabinsk event, 2013-02-15 03:20:34 UTC.  Top panel: color coded probability at each random location.  Bottom panel: contour map obtained after interpolating probabilities.  The black star shows the position of Chelyabinsk on the maps, and the impact probability for this site on the color bar of the bottom panel.  \hll{Red and blue circles} in the bottom panel indicate, respectively, the projection of the apex and antapex on the Earth's surface.  Sites located on the dashed line face in a direction perpendicular to apex-antapex direction (apex equator, $\theta\sub{apex}^{\rm geo}=90\deg$).  Continuous solid line is the projection of the ecliptic on Earth's surface. Both circumferences intersect at points where local time is 12 \hll{noon} (to the east of apex, at this time in the northern coast of Australia) and 12 \hl{midnight} (to the west of apex, at this time in front of the Guyana coast).\vspace{0.2cm}}
\label{fig:ProbabilityMapChelyabinsk}
\end{figure*}

Relative impact probability calculated with GRT has a non-trivial geographical distribution. As expected from our previous experiments, sites close to  {the greographic} apex and antapex have low and large probabilities {, respectively}.  The largest impact probabilities at the time of Chelyabinsk impact, however, were not at the geographic antapex nor around the Chelyabinsk region.   {They were} around the southern coast of South America,  {a region that at the time of the event} was almost at the geographic apex equator ($\theta\sub{apex}^{\rm geo}\sim 90\deg$); the local time in this region was close to 12 \hll{midnight}.  

Sites located close to the geographic apex equator have probabilities systematically larger than the rest of the world (with the exception of the antapex regions).  These probabilities are dependent on geographic apex longitude $\lambda\sub{apex}\sup{geo}$ ($\lambda\sub{apex}\sup{geo}=0$ at the subsolar point).  Points close to the apex equator in the day side of the Earth have lower impact probabilities than points in the antisolar region. This asymmetry arises from a dynamical effect, namely particles thrown from the day side end up in orbits with lower perihelion distances where the density of NEOs is lower.  The asymmetry could explain the day-night (AM/PM) asymmetry, identified in other works \hll{(see eg. \citealt{Gallant2009})}.

The previous calculation was repeated in the case of the Tunguska and the 1963 events.  The resulting contour maps are shown in \autoref{fig:ProbabilityMapsOther}. These maps confirm our previous conclusions. The geographical distribution of impacts has a significant dipolar asymmetry in the apex-antapex directions. Sites closer to the geographic antapex have slightly larger impact probabilities than the rest of the planet.  The regions with the largest probabilities are closer to the apex equator but such that $\theta\sub{apex}^{\rm geo}\lesssim 90\deg$.  AM/PM asymmetries are also noticeable at the time of Tunguska and 1963 impacts  (extended red areas in the maps).  

\begin{figure*}  
  \centering
  \vspace{0.2cm}
   \includegraphics[width=115mm]{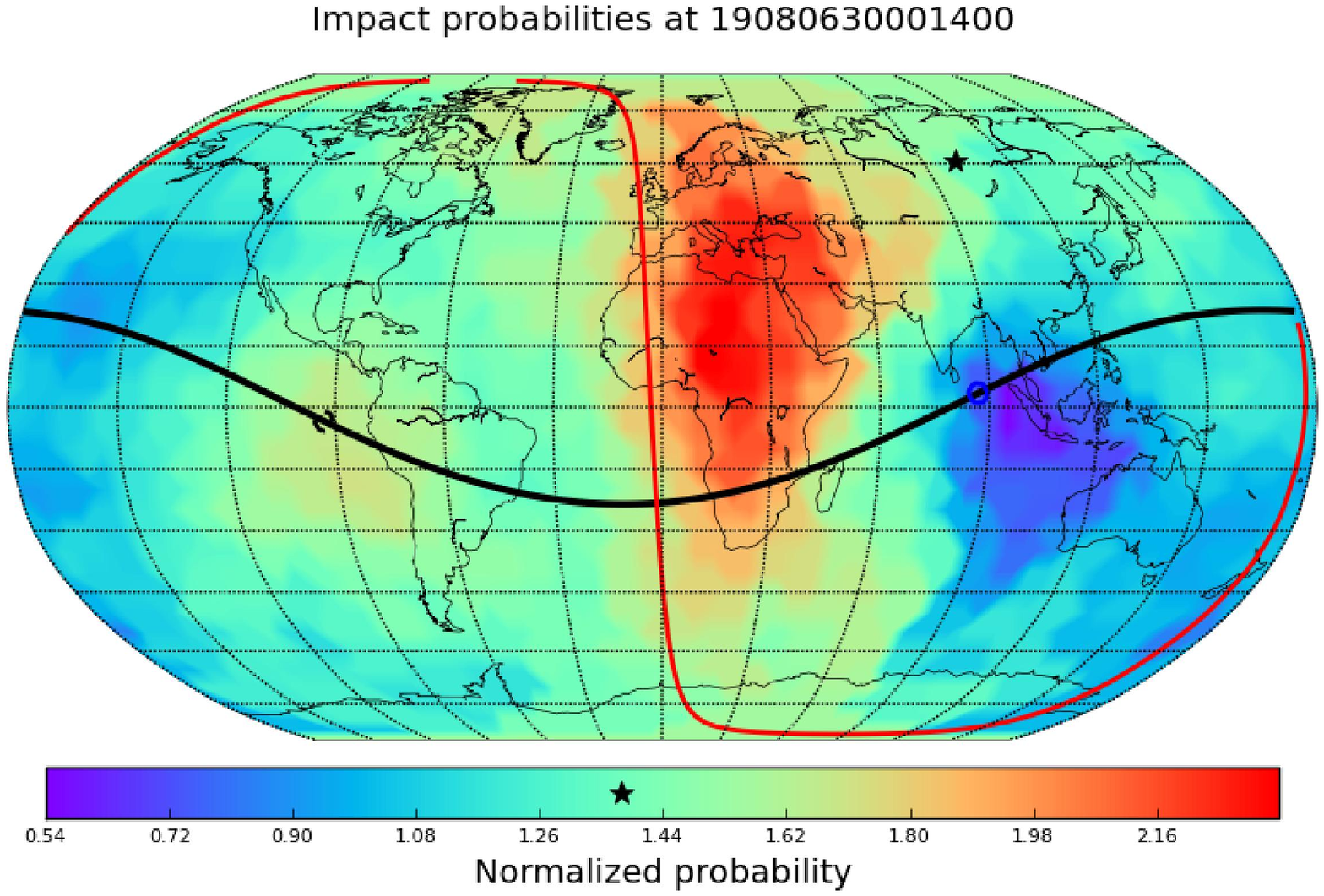}\\\vspace{0.2cm}
   \includegraphics[width=115mm]{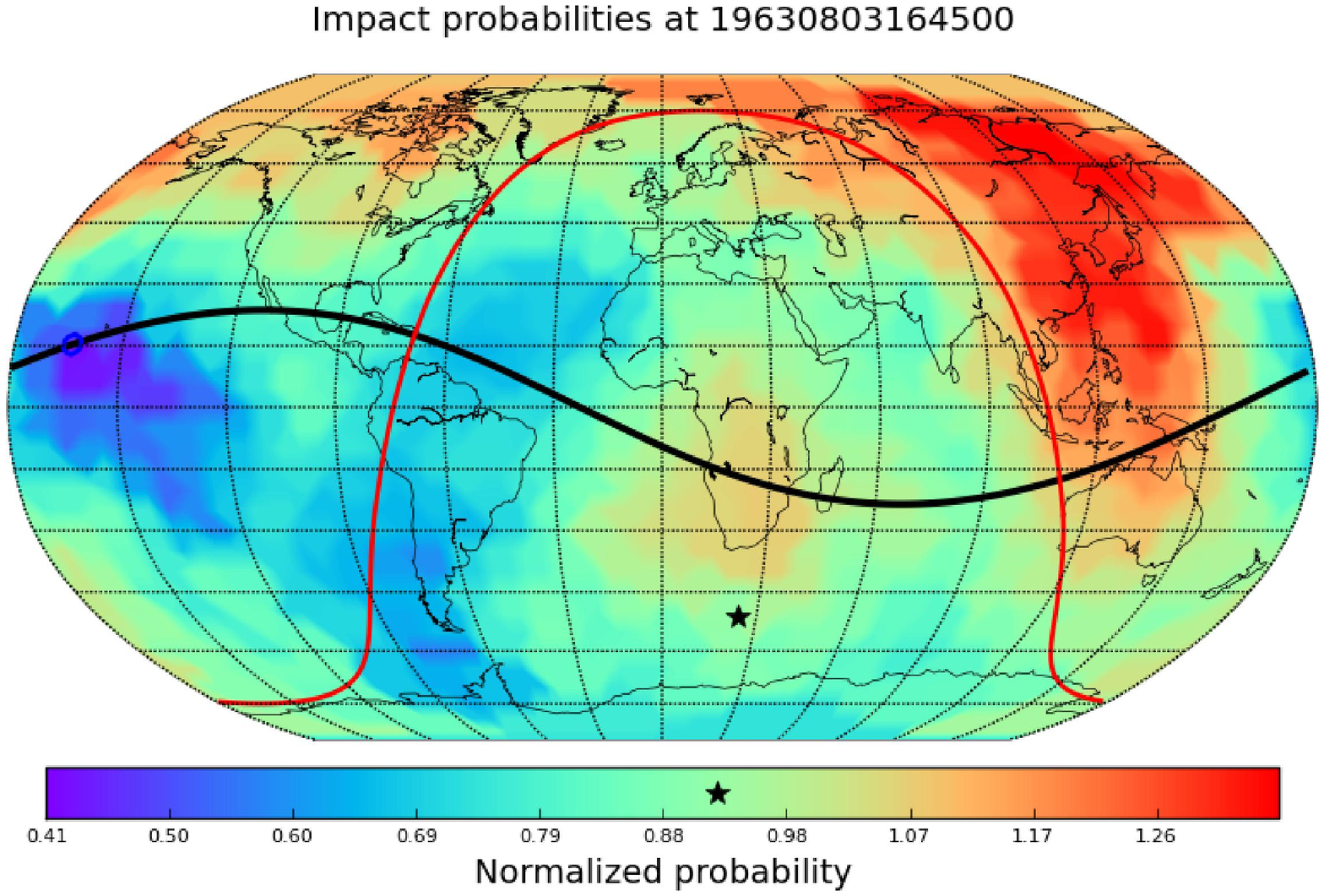}\\\vspace{0.2cm}
   \scriptsize
  \caption{Contour maps of impact probability across the Earth's surface.  Top panel: at the date and time of the Tunguska event (June 30th, 1908 00:14:00 UTC). Bottom panel: at the date and time of the 1963 event identified by \citep{Revelle1997} (Aug. 3rd, 16:45:00 UTC).  Black stars in both cases indicate the geographical position and probability of the impact events.\vspace{0.2cm}}
\label{fig:ProbabilityMapsOther}
\end{figure*}

\subsection{Impact velocity distribution}
\label{subsec:impact_velocity}

 When calculating the impact probability $P\sub{imp}$ for a given geographical location, we need to compute first, initial condition probabilities $P^{*}\sub{ini}(A,z,\vimp;t)$. Besides the natural role of this quantity in calculating the total impact probability, it could also provide other interesting statistical information.

Thus, for instance, an integration over all possible impact velocities at a given direction in the sky $(A_i,h_i)$, namely,

$$
\sum_j P^{*}\sub{ini}(A_i,h_i,v\sub{imp,j};t),
$$

will provide the distribution of incoming directions. In other words we can determine which regions in the sky would be more prone to impacts in a given location and time.

If otherwise we integrate $P\sup{*}\sub{ini}(A,z,\vimp;t)$ over all the possible incoming directions, the probability distribution of impact speeds can be obtained. Since the distribution of impact speeds is known from meteor and fireball observations, this calculation constitute an interesting way to check the GRT method.

\autoref{fig:VelocityDistribution} shows the reconstructed distribution of impact speeds as obtained from the simulations performed in \autoref{subsec:discrete_locations}.  

We notice that independent of the fact that a regularly spaced set of velocities were used to perform the GRT analysis, a non uniform distribution of impact speeds arises from impact probabilities calculated with the methods here.  Moreover, the distribution closely matches the observed distribution of fireballs.  

As expected, points located in the geographic apex (Madagascar in this case) have impact speeds larger than in the geographic antapex (Hawaii).  These particularities are washed out when averaging over a large sample of meteor and impacts happening in different locations and times.

\begin{figure}  
  \centering
  \vspace{0.2cm}
   \includegraphics[scale=0.45]{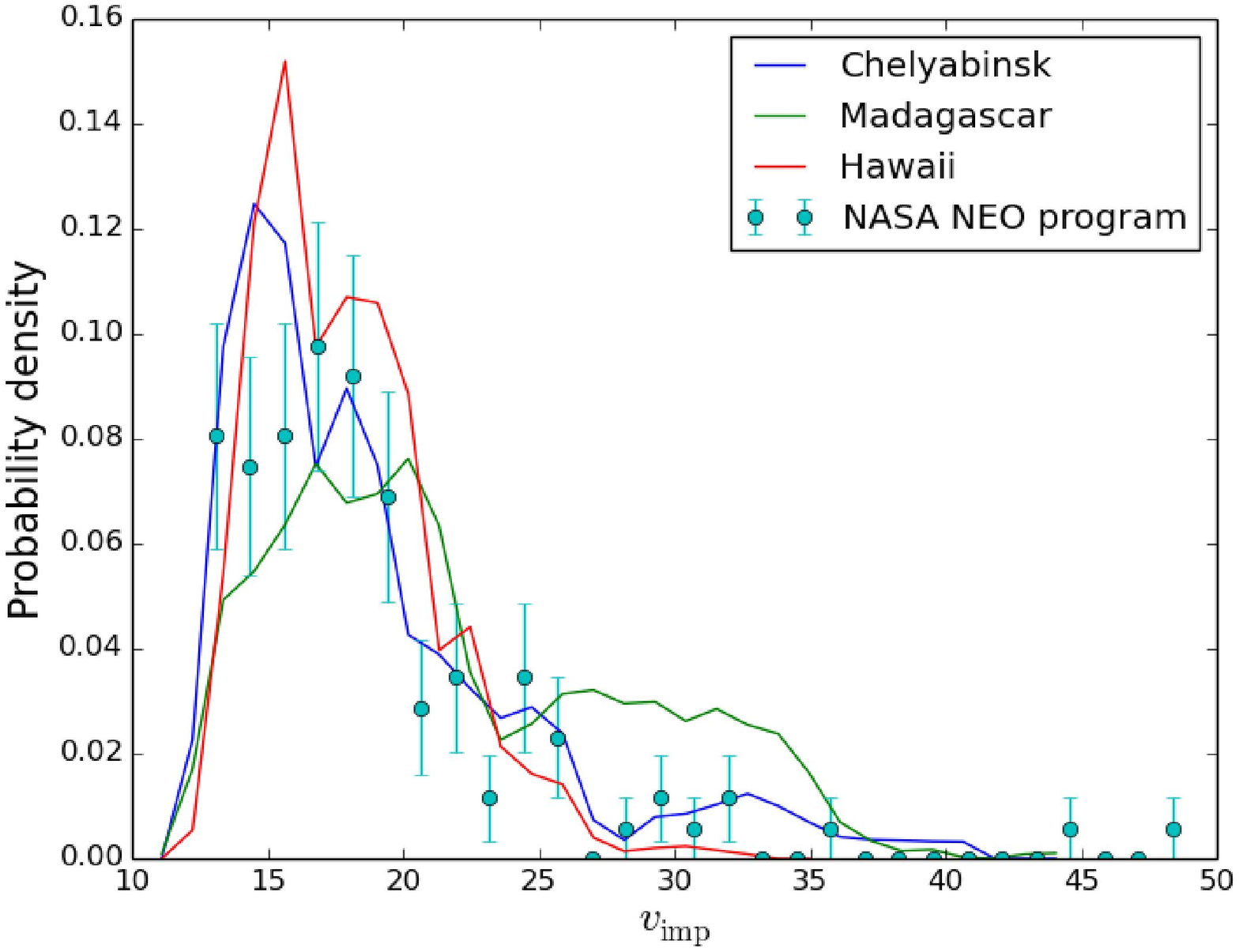}\\\vspace{0.2cm}
  \caption{Velocity distribution as reconstructed from the initial condition probability distribution $P\sup{*}\sub{ini}(A,z,\vimp;t)$ at the time of the Chelyabinsk event. For comparison we include the impact speed distribution calculated from the NASA bolide database \hll{\url{http://neo.jpl.nasa.gov/fireballs}}.\vspace{0.2cm}}
\label{fig:VelocityDistribution}
\end{figure}

\section{Discussion}
\label{sec:discussion}

The maps in \autoref{fig:ProbabilityMapChelyabinsk} and \autoref{fig:ProbabilityMapsOther} represent the first documented attempt at predicting theoretically the instantaneous geographical distribution of impact probabilities on Earth. Although their resolution is still poor ($\sim 800$ km),  the method can be in general applied to study the impact probability at smaller geographical regions, without increasing considerably the computational cost.

For making each map,  $>$200,000 test particles were thrown from all around the Earth.  As expected, not all of them contribute significantly to the calculation of impact probabilities.  In some cases a test particle thrown from a given site, collided again with Earth or ends up at the surface of the Sun. In other cases the resulting heliocentric orbits were hyperbolic or retrograde.  On average  $\sim 60\%$ of the test particles used by a GRT analysis actually contribute to compute impact probability (efficiency can be increased if instead of using a uniform distribution of impact velocities we use a distribution based on meteor observations). However, in contrast to forward-integration methods, in GRT even the lost particles are important to weight the total impact probability of a given site.   GRT represents a gain in efficiency of several orders of magnitudes with respect to the usual forward-integration methods.  

Counterintuitively and according to GRT, the geographic anteapex regions have larger impact probability than those closer to geographic apex. This effect can be explained by the fact that to have an impact on the anteapex with a given impact speed $\vimp=\sqrt{\vinf^2+\vesc^2}$, the ecliptic velocity of the impactor needs to be $\sim(\vinf+\vearth)>\vearth$.  In order to have this velocity the impactor should be in an eccentric orbit with perihelion distance close to 1 au.  Since the density of NEOs in the configuration space is large around $q\sim 1$ au, $e\sim 0.5$ (see \autoref{fig:NEODistribution}), this condition will be more probable.  On the other hand if an impact occurs in the geographic apex with the same impact velocity $\vimp$, the ecliptic velocity of the impactor will be $\sim(\vearth-\vinf)<\vearth$. Such heliocentric velocities at $r=1$ au correspond to highly eccentric orbits with low perihelion distances where the density of NEOs is much lower.  \hll{It is important to stress here, that the lack objects with low perihelion distances, may arise from the fact that these objects are hard to be detected at opposition and not from their actual distribution.  If this is the case, the apex-anteapex asymmetry could be an ``observational'' artifact rather than a physical effect (see \autoref{sec:conclusions}).}

The fact that none of the largest impacts studied here, occurred in regions with the largest probabilities as predicted by GRT, may imply two things: (1) the probability distributions calculated here are biased or simply wrong or (2) three events does not constitute enough statistical evidence to test the validity of our predictions.   Impact of meteoroids is a stochastic process.  Even if a geographical site is more prone to an impact at a given time, it does not imply that an actual impact will happen at that location.

This paper begins by raising the question of why the Chelyabinsk and Tunguska events happened just 2,400 km apart. Although we did not find a definitive answer to this ``puzzle'' (if there is actually one), several interesting facts arose during our investigation. The first one is that at the latitude where both events happened $\phi\sim 60\deg$, the apex colatitude varies during the year between $\theta\sub{apex}^{\rm geo}\sim 30\deg$ and $\theta\sub{apex}^{\rm geo}\sim150\deg$.  We expect that during the greater part of the day, both Tunguska and Chelyabinsk regions are at apex colatitudes where large impact probabilities exist.  But this condition is not restricted to those sites. In the northern hemisphere Norway, Canada and Alaska, are in continental areas with similar latitudes. In the southern hemisphere, however, geographical locations with similar conditions are in the middle of the ocean (Indian, Altantic and Pacific). Another interesting common condition between the Tunguska and Chelyabinsk regions at the date and time of the impacts is that both happen around the same universal time ($0-3$ UTC). Accordingly, both sites were around the same geographical hemisphere as the apex (see maps in \autoref{fig:ProbabilityMapChelyabinsk} and \autoref{fig:ProbabilityMapsOther}) and above the apex equator.  These are precisely the regions where according to GRT the largest probabilities are found.

\section{Summary and conclusions}
\label{sec:conclusions}

We presented in this paper a novel method to calculate the relative probability of an asteroid or meteoroid impact on the surface of a Solar System body.  We have emphasized on the case of impacts on Earth but the method can be extended to other planetary objects.  

The method, called Gravitational Ray Tracing or GRT, is inspired by an analogue simulation technique used in computer graphics to render complex visual scenes.  We have adopted many of the features of the optical method to develop GRT, and even several complimentary numerical techniques (eg. sampling algorithms). 

GRT relies on our capability to know a priori the distribution of Near Earth Objects (NEOs) in configuration space.

We confirmed our initial hypothesis that at a given time not all points on Earth are equally prone to impacts (geographical instantaneous distribution is not uniform).  Geographical areas pointing in the same direction of the apex projection on Earth's surface (geographic apex) have lower impact probabilities than those located in the anteapex direction. The largest probabilities predicted by our method are however, localized on extensive geographical areas around to geographic apex equator  (locations located in directions perpendicular to the apex-antapex direction).  Our results predict an AM/PM asymmetry, although we did not quantify its size.

No obvious relationship among the geographical areas of the Chelyabinsk and Tunguska events were discovered with GRT. However, several interesting common features of their locations were identified.  Chelyabinsk and Tunguska areas are located during most part of the year at low apex latitudes  (large geographic apex colatitudes), where impact fluxes are larger.  Moreover, the impact events at both sites, happened around the same time of the day and the AM/PM asymmetry could increase the chances for an impact on these locations.

We calculate the impact probability with GRT at the locations and times of 394 fireballs of the NASA's bolide database. Although probability values does not follow the expected distribution we hope that improving the flux correcting function may allow us to improve the \hl{matching between the model and the observed impact distribution}.

GRT reproduces well the observed distribution of meteoroid impact \hl{speeds}.  Besides the fact that we start with a uniform distribution of impact speeds for the test particles, not all the speeds had the same probability of producing a viable asymptotic orbit.  Moreover, GRT is capable at predicting subtle differences between the distribution of impact speeds at different locations, dates and times which is normally hard to achieve using only fireball observations.

\hl{This paper did not aim at exploring exhaustively the applications of the GRT method.  Neither was our goal to declare that the results presented here are definitive or general. Our main goal was to describe and apply the method to realistic cases, opening the door to future improvements. There are multiple aspects of the method that need to be improved.  The effect that the flux correcting function have in the relative probability must be explored in detail.  The dependency of this function on $\lambda\sub{apex}$ should also be included and its effect on the resulting relative probability, analyzed.  Larger resolution analyses should be performed to see the effect that ``small-numbers statistics'' could have in the results. A very important improvement involves using a properly debiased population of NEOs.  For instance, the observed strong apex-antapex asymmetry obtained here, could arise, not from the population of NEOs and their dynamics relative to Earth, but from the fact that objects having aphelia close to 1 AU cannot be observed at opposition. Also a proper consideration of the effect that a different spatial distribution of small (H>20) objects with respect to the larger ones, against which we compare our rays, should be investigated. Probability calculation and normalization could also be improved with a better approximation to the continuous probability in \autoref{eq:ImpactProbability}. Last but not least, a proper effort to fit using GRT the observed spatio-temporal distribution of fireballs, should be undertaken. Interestingly, the latter effort could even contribute to a better understanding of the NEOs population by suggesting the absence or presence of objects required to explain the observations.}

This is just the beginning of a larger effort to asses the impact risk on Earth using backward-integration methods.  If successful \hl{in the long run}, this effort could open the door to better understanding the chances of moderate size impacts that could be used for decision making and other mitigation purposes.

\section*{Reproducibility}

In pursuance of making these results public and reproducible, Zuluaga J. I. developed, tested and documented a public package, {\tt GravRay}, that can be freely downloaded from a {\tt GitHub} repository: \url{http://github.com/seap-udea/GravRay} (branch {\tt GravRay\_alpha2}). The package includes several open source third party software (NASA NAIF SPICE and GNU Scientific Library) and several databases (Minor Planet Center database, NASA fireball and bolide database and NASA Small Bodies database). The author of the package appreciate any feedback or bug report.

\section*{Acknowledgements}

We have used NASA's ADS Bibliographic Services. Most of the computations that made possible this work were performed with NASA NAIF SPICE Software (\citealt{Acton1996} and Jon D. Giorgini), {\tt Python 2.7} and their related tools and libraries, {\tt iPython} \citep{Perez2007}, {\tt Matplotlib} \citep{Hunter2007}, {\tt scipy} and {\tt numpy} \citep{Van2011}. \hl{Special thanks to the reviewers that through their criticism, significantly contributed to improve the paper}. \hl{In particular,} we thank Steve Chesley for \hll{his} in-depth \hl{initial} review of this work; \hll{his} comments and observations \hl{contributed} to improve significantly the quality of the manuscript with respect to its initial versions. We appreciate the assistance of Jon D. \hll{Giorgini}, senior analyst of the JPL Solar System Dynamics Group. We also thanks to Prof. Ignacio Ferrin and Prof. David Asher for useful comments and insightful discussions about this topic.  \hl{We also thank to the organizers of Meteoroid 2016 conference, for allowing us to present the initial versions of this work to the meteor and asteroid scientific community}.  This work is supported by Vicerrectoria de Docencia-UdeA and the {\it Estrategia de Sostenibilidad 2014-2015 de la Universidad de Antioquia}.  MS is supported by Colciencias, {\it Doctorado Nacional - 647} program.   {We thanks Universidad Pontificia Bolivariana (UPB-Medellin) and DELL for allows us to run most of our simulations in their data centers}.



\bsp	
\label{lastpage}
\end{document}